\documentclass[12pt]{article}

\usepackage{mdwlist}
\usepackage{enumerate}

\usepackage{float}
\usepackage{cite}
\usepackage{amsfonts}
\usepackage{amssymb}
\usepackage{amsmath}
\usepackage{xcolor}
\usepackage{mathtools}

\usepackage{graphicx}

\usepackage{titlesec}

\titlespacing*{\section}
{0pt}{0.5ex plus .2ex minus .2ex}{0.5ex plus .2ex}
\titlespacing*{\subsection}
{0pt}{0.5ex plus .2ex minus .2ex}{0.5ex plus .2ex}

\def\comp{{\rm C}\llap{\vrule height7.1pt width1pt depth-.4pt\phantom t}}
\def\gtwid{\mathrel{\raise.3ex\hbox{$>$\kern-.75em\lower1ex\hbox{$\sim$}}}}
\def\ltwid{\mathrel{\raise.3ex\hbox{$<$\kern-.75em\lower1ex\hbox{$\sim$}}}}
\def\square{\kern1pt\vbox{\hrule height 1.2pt\hbox{\vrule width 1.2pt\hskip 3pt
   \vbox{\vskip 6pt}\hskip 3pt\vrule width 0.6pt}\hrule height 0.6pt}\kern1pt}

\begin{document}

\begin{titlepage}

\begin{flushright}
UFIFT-QG-25-01
\end{flushright}

\vskip 3cm

\begin{center}
{\bf Resummations for Inflationary Quantum Gravity}
\end{center}

\vskip 1cm

\begin{center}
R. P. Woodard$^{\dagger}$
\end{center}

\vskip 0.5cm

\begin{center}
\it{Department of Physics, University of Florida,\\
Gainesville, FL 32611, UNITED STATES}
\end{center}

\vspace{0.5cm}

\begin{center}
ABSTRACT
\end{center}
The continual production of gravitons during inflation endows loop 
corrections with secular logarithms which grow nonperturbatively large 
during a prolonged period of inflation. The physics behind these 
effects is reviewed, along with a catalog of the examples which have 
so far been found. Resummation can be accomplished by combining a
variant of Starobinsky's stochastic formalism with a variant of the
renormalization group. The issue of gauge independence is also
addressed.

\begin{flushleft}
\end{flushleft}

\vskip 3cm

\begin{flushleft}
$^{\dagger}$ e-mail: woodard@phys.ufl.edu
\end{flushleft}

\end{titlepage}

\section{Introduction}

The background geometry of cosmology can be represented in terms of a scale
factor $a(t)$, Hubble parameter $H(t)$ and first slow roll parameter 
$\epsilon(t)$,
\begin{equation}
ds^2 = -c^2 dt^2 + a^2(t) d\vec{x} \!\cdot\! d\vec{x} \qquad , \qquad H(t)
\equiv \tfrac{\dot{a}}{a} \quad , \quad \epsilon(t) \equiv -\tfrac{\dot{H}}{H^2}
\; . \label{background}
\end{equation}
Inflation is defined as accelerated expansion, which means $H(t) > 0$ and
$0 \leq \epsilon(t) < 1$. It is occurring now \cite{DES:2021wwk,
Kamionkowski:2022pkx} and a vast amount of data supports the belief that it 
also took place during the very early universe \cite{Planck:2018vyg,
Geshnizjani:2011dk,Kinney:2012zbd}. The accelerated expansion of primordial 
inflation lead to the copious production of light, minimally coupled scalars 
and gravitons, which is what caused the scalar and tensor power spectra 
\cite{Mukhanov:1981xt,Starobinsky:1979ty}.

At some level the scalars and gravitons produced during inflation must interact 
with themselves and with other particles. Scalar effects are simpler to compute, 
but they are highly model dependent. In contrast, graviton effects require much 
greater effort, but they are completely generic as long as general relativity 
can be regarded as the low energy effective field theory of gravity in the sense 
of Donoghue \cite{Donoghue:1993eb,Donoghue:1994dn,Donoghue:2017ovt}. This paper
is devoted to explaining recent progress in the elucidation of these generic 
graviton effects.

Loops of inflationary gravitons tend to grow in space and time because more and
more gravitons are produced as inflation progresses \cite{Miao:2006gj,
Glavan:2013jca,Wang:2014tza,Glavan:2021adm,Tan:2021lza,Tan:2022xpn}. For 
example, de Sitter ($\epsilon = 0$) gravitons change the Coulomb potential
\cite{Glavan:2013jca} and the electric field strength of plane wave photons 
\cite{Wang:2014tza} to,
\begin{eqnarray}
\Phi(t,r) &\!\!\! = \!\!\!& \tfrac{Q}{4\pi \varepsilon_0 a(t) r} \Bigl\{1
+ \tfrac{2\hbar G}{3 \pi c^3 a^2(t) r^2} + \tfrac{2 \hbar G H^2}{\pi c^5} 
\ln[\tfrac{a(t) H r}{c}] + \dots \Bigr\} \; , \qquad \label{Coulomb} \\
F_{0i}(t,\vec{x}) &\!\!\! = \!\!\!& F_{0i}^{\rm tree}(t,\vec{x}) \Bigl\{1 +
\tfrac{2 \hbar G H^2}{\pi c^5} \ln[a(t)] + \dots \Bigr\} \; . \qquad 
\label{photon}
\end{eqnarray}
The fractional $2 \hbar G/3 \pi c^3 a^2 r^2$ correction to (\ref{Coulomb}) 
is the de Sitter descendant of an effect long known from work on flat space 
background \cite{Radkowski:1970}. The new effects due to inflationary 
particle production are proportional to $\hbar G H^2/c^5$.

During a prolonged period of inflation the growth of $\ln[a(t)] = H t$ must 
eventually make these contributions of order one, no matter how small the 
dimensionless loop-counting parameter $\hbar G H^2/c^5$. This does not 
necessarily lead to large effects because the leading higher loop contributions 
are also of order one, but it does mean that perturbation theory has broken 
down. A nonperturbative resummation scheme is required to work out what happens 
at late times. The search for such a scheme has been long and confusing because
large loop corrections derive from two distinct sources, the resummation of
which each requires its own technique. This work describes the synthesis of the 
two techniques which leads to a complete resummation.

This article has six sections of which this Introduction is the first. In all
subsequent sections the standard $\hbar = c = 1$ units of particle physics are
employed, and $\kappa^2 \equiv 16 \pi G$ is used in place of Newton's constant. 
The two sources of large loop corrections are ``tail'' terms of propagators and 
the mismatch between primitive divergences and the counterterms which absorb 
them. Section 2 explains this and presents a catalog of the dimensionally 
regulated \cite{tHooft:1972tcz,Bollini:1972ui} and fully BPHZ renormalized 
(Bogoliubov, Parasiuk \cite{Bogoliubov:1957gp}, Hepp \cite{Hepp:1966eg} and 
Zimmermann \cite{Zimmermann:1968mu,Zimmermann:1969jj}) results which have so 
far been derived on de Sitter background. Section 3 describes the stochastic 
formalism which facilitates the resummation of large tail corrections. The 
resummation of the other kind of large correction is presented in section 4. 
These resummation schemes are implemented in a fixed gauge which raises the 
issue of gauge dependence. Section 5 discusses how to resolve that. Conclusions
and open problems are presented in section 6. 

\section{Large Logarithms from Inflationary QFT}

The point of this section is to describe the various large logarithms which
appear in loop corrections on de Sitter background, both to matter from 
inflationary gravitons and to gravity from inflationary matter. We begin by 
identifying the two sources which produce them. Then we discuss the graviton 
gauge which has been used for all but one of the existing computations. The 
section closes with a review of results.

\subsection{Why Large Loop Corrections Happen}

As mentioned before, there are two sources of large logarithms. The first
comes from what DeWitt and Brehme termed the ``tail'' part of the graviton
propagator \cite{DeWitt:1960fc}. To understand what this means, recall that
dynamical gravitons have the same mode functions as those of massless,
minimally coupled scalars \cite{Lifshitz:1945du}. Hence the propagators of
massless, minimally coupled scalars and gravitons must agree. On 
4-dimensional de Sitter this propagator is,
\begin{equation}
D=4 \qquad \Longrightarrow \qquad i\Delta_A(x;x') = \tfrac1{4\pi^2} 
\tfrac{1}{a a' \Delta x^2} - \tfrac{H^2}{8\pi^2} \ln[\tfrac14 H^2 
\Delta x^2] \; , \label{4Dtail}
\end{equation}
where $\Delta x^2 \equiv \eta_{\mu\nu} (x - x')^{\mu} (x - x')^{\nu}$ is 
the interval in conformal coordinates $x^{\mu} \equiv (\eta,\vec{x})$ with
$d\eta = dt/a(t)$. The ``tail'' in this expression is the logarithm. The 
general $D$ propagator employed for dimensional regularization is 
\cite{Onemli:2002hr,Onemli:2004mb},
\begin{eqnarray}
\lefteqn{i \Delta_A(x;x') = \tfrac{\Gamma(\frac{D}{2} - 1)}{4 \pi^{\frac{D}2}}
\tfrac{1}{[a a' \Delta x^2]^{\frac{D}2 - 1}} } \nonumber \\
& & \hspace{2cm} + \tfrac{\Gamma(\frac{D}2+1)}{8 \pi^{\frac{D}2} (D-4)} 
\tfrac{H^2}{[a a' \Delta x^2]^{\frac{D}2 - 2}} - k [\pi \cot(\tfrac{D \pi}{2}) 
- \ln(a a') ] \qquad \nonumber \\
& & \hspace{3cm} + \tfrac{\Gamma(\frac{D}2+2)}{64 \pi^{\frac{D}2} (D-6)} 
\tfrac{H^4 a a' \Delta x^2}{[a a' \Delta x^2]^{\frac{D}2 - 2}} + (\tfrac{D-1}{2D})
k H^2 a a' \Delta x^2 + \dots \; , \qquad \label{Adef}
\end{eqnarray}
where the constant $k$ is,
\begin{equation}
k \equiv \tfrac{H^{D-2}}{(4\pi)^{\frac{D}2}} \tfrac{\Gamma(D-1)}{
\Gamma(\frac{D}2)} \; . \label{kdef}
\end{equation}
The tail of the general $D$ propagator is the second line of expression 
(\ref{Adef}). This tail is responsible for two of the coincidence
limits employed later,
\begin{equation}
i\Delta_A(x;x')_{x'=x} = - k [\pi \cot(\tfrac{D \pi}{2}) 
- 2 \ln(a) ] \quad , \quad \partial_{\mu} i\Delta_A(x;x')_{x'=x} = a H k 
\delta^{0}_{~\mu} \; . \label{2coinc}
\end{equation}
A final coincidence limit comes from the last term in expression 
(\ref{Adef}),
\begin{equation}
\partial_{\mu} \partial'_{\nu} i \Delta_A(x;x')_{x' = x} = -(\tfrac{D-1}{D})
k H^2 g_{\mu\nu} \; . \label{3rdcoinc}
\end{equation}

The second source of large logarithms is through renormalization. In 
dimensional regularization it turns out that primitive divergences never
involve $D$-dependent powers of the scale factor. This is because vertices
contribute a factor of $a^D$ from the measure factor, whereas each propagator
from $x^{\mu}$ to ${x'}^{\mu}$ contributes a factor of $(a a')^{1-\frac{D}2}$
times integer powers of $a a'$, as shown in expression (\ref{Adef}). Hence 
the $D$-dependent scale factors cancel when two propagators join two vertices. 
On the other hand, counterterms inherit an uncanceled factor of $a^D$ from the 
measure, which results in a large logarithm after subtraction 
\cite{Miao:2021gic},
\begin{equation}
\tfrac{(2H)^{D-4}}{D - 4} - \tfrac{(\mu a)^{D-4}}{D - 4} = 
-\ln(\tfrac{\mu a}{2H}) + O(D \!-\! 4) \; . \label{RGsource}
\end{equation}

\subsection{Gauge Fixing on de Sitter}

The Lagrangian for $D$-dimensional gravity with a positive cosmological
constant is,
\begin{equation}
\mathcal{L}_{\rm Einstein} = \tfrac{[R - (D-2) \Lambda]}{16 \pi G} \; .
\label{Einstein}
\end{equation}
The maximally symmetric solution to this theory is de Sitter, with a zero 
first slow roll parameter and constant Hubble parameter obeying $\Lambda = 
(D-1) H^2$. The geometry takes the form (\ref{background}) with scale
factor $a(t) = e^{H t}$, and we shall often employ conformal time $\eta
= -e^{-H t}/H$ so that the background metric takes the form $a^2 
\eta_{\mu\nu}$. The graviton field $h_{\mu\nu}(x)$ is defined by
conformally transforming the full metric,
\begin{equation}
g_{\mu\nu} \equiv a^2 \widetilde{g}_{\mu\nu} \equiv a^2 [\eta_{\mu\nu} 
+ \kappa h_{\mu\nu}] \qquad , \qquad \kappa^2 \equiv 16 \pi G \; .
\label{hdef}
\end{equation}
Graviton indices are raised and lowered using the Minkowski metric,
$h^{\mu}_{~\nu} \equiv \eta^{\mu\rho} h_{\rho\nu}$, and its trace is
$h \equiv \eta^{\mu\nu} h_{\mu\nu}$.

Because the spatial plane wave mode functions for dynamical gravitons 
are identical to those of the massless, minimally coupled scalar 
\cite{Lifshitz:1945du}, the propagator for which breaks de Sitter 
invariance \cite{Allen:1987tz}, one would think that the graviton
propagator must also break de Sitter invariance. However, this has 
long been a contentious issue because the imposition of certain
gauges (often with analytic continuation from Euclidean space) does 
permit a de Sitter invariant solution for the propagator
\cite{Allen:1986ta,Allen:1986tt,Hawking:2000ee,Higuchi:2001uv,
Higuchi:2002sc,Morrison:2013rqa}, except for an infinite class of 
discrete choices \cite{Antoniadis:1986sb,Folacci:1992xc,Folacci:1996dv}. 
Three key insights seem to resolve the controversy in favor of de Sitter 
breaking:
\begin{itemize*}
\item{The linearization instability of de Sitter precludes the
addition of any de Sitter invariant gauge fixing term to the 
action \cite{Miao:2009hb}, although one can still impose de Sitter
invariant gauges as strong operator equations.}
\item{The analytic continuation techniques which are invoked to 
pass from Euclidean de Sitter fail to detect power law infrared
divergences \cite{Kiritsis:1994ta,Janssen:2008px,Miao:2013isa}
which obstruct the continuation. The problematic discrete choices 
correspond to the appearance of logarithmic infrared divergences
which do show up in analytic continuation.}
\item{Not all solutions to the propagator equation represent true
propagators in the sense of being the expectation value of the 
time-ordered product of two fields \cite{Janssen:2008px}. The
results for exact de Sitter invariant gauges in Lorentzian signature
which are true propagators all break de Sitter invariance 
\cite{Miao:2011fc,Mora:2012zi}.}
\end{itemize*}

Because the graviton propagator must in any case break de Sitter
invariance, there is no reason to avoid imposing a noninvariant 
gauge, and doing so can result in a substantially simpler 
propagator. The simplest choice consists of adding the gauge 
fixing functional\cite{Tsamis:1992xa,Woodard:2004ut},
\begin{equation}
\mathcal{L}_{GF} = -\tfrac12 a^{D-2} \eta^{\mu\nu} F_{\mu} F_{\nu}
\quad , \quad F_{\mu} = \eta^{\rho\sigma} \Bigl[ h_{\mu\rho , \sigma}
- \tfrac12 h_{\rho\sigma , \mu} + (D\!-\!2) a H h_{\mu\rho} 
\delta^0_{~\sigma} \Bigr] . \label{LGF}
\end{equation}
In this gauge the graviton and ghost propagators are,
\begin{eqnarray}
i[\mbox{}_{\mu\nu} \Delta_{\rho\sigma}](x;x') &\!\!\! = \!\!\!& 
\sum_{I=A,B,C} [\mbox{}_{\mu\nu} T^I_{\rho\sigma}] \!\times\! 
i\Delta_I(x;x') \; , \qquad \label{gravprop} \\
i[\mbox{}_{\mu} \Delta_{\rho}](x;x') &\!\!\! = \!\!\!& 
\overline{\eta}_{\mu\rho} \!\times\! i\Delta_A(x;x') - \delta^0_{~\mu}
\delta^0_{~\rho} \!\times\! i\Delta_B(x;x') \; . \qquad \label{ghostprop}
\end{eqnarray}
Here $\overline{\eta}_{\mu\nu} \equiv \eta_{\mu\nu} + \delta^0_{~\mu}
\delta^0_{~\nu}$ is the purely spatial part of the Minkowski metric.

One of the things that makes the propagator in this gauge so simple is
that the tensor factors are constant,
\begin{eqnarray}
[\mbox{}_{\mu\nu} T^A_{\rho\sigma}] & = & 2 \overline{\eta}_{\mu (\rho}
\overline{\eta}_{\sigma) \nu} - \tfrac{2}{D-3} \overline{\eta}_{\mu\nu}
\overline{\eta}_{\rho\sigma} \qquad , \qquad [\mbox{}_{\mu\nu} 
T^B_{\rho\sigma}] = -4 \delta^0_{~(\mu} \overline{\eta}_{\nu) (\rho}
\delta^0_{~\sigma)} \; , \qquad \label{TATB} \\
\, [\mbox{}_{\mu\nu} T^C_{\rho\sigma}] & = & \tfrac{2}{(D-2) (D-3)} [(D\!-\!3)
\delta^0_{~\mu} \delta^0_{~\nu} \!+\! \overline{\eta}_{\mu\nu}] [(D\!-\!3)
\delta^0_{~\rho} \delta^0_{~\sigma} \!+\! \overline{\eta}_{\rho\sigma}]
\; . \qquad \label{TC}
\end{eqnarray}
The $A$ propagator is expression (\ref{Adef}); the $B$ and $C$ propagators 
are,
\begin{eqnarray}
\lefteqn{i\Delta_B = \tfrac{\Gamma(\frac{D}2 - 1)}{4 \pi^{\frac{D}2}} 
\tfrac{1}{[a a' \Delta x^2]^{\frac{D}2 - 1}} + \tfrac{\Gamma(\frac{D}2)}{16 
\pi^{\frac{D}2}} \tfrac{H^2}{[a a' \Delta x^2]^{\frac{D}2 - 2}} - 
\tfrac{k}{D-2} } \nonumber \\
& & \hspace{4cm} + \tfrac{\Gamma(\frac{D}2 + 1)}{128 \pi^{\frac{D}2}}
\tfrac{H^4 a a' \Delta x^2}{[a a' \Delta x^2]^{\frac{D}2 - 2}} - \tfrac{k}{2D} 
H^2 a a' \Delta x^2 + \dots \; , \qquad \label{Bdef} \\
\lefteqn{i\Delta_C = \tfrac{\Gamma(\frac{D}2 - 1)}{4 \pi^{\frac{D}2}} 
\tfrac{1}{[a a' \Delta x^2]^{\frac{D}2 - 1}} + \tfrac{(\frac{D}2 - 3)
\Gamma(\frac{D}2 - 1)}{16 \pi^{\frac{D}2}} \tfrac{H^2}{[a a' \Delta x^2]^{
\frac{D}2 - 2}} + \tfrac{k}{(D-2) (D-3)} } \nonumber \\
& & \hspace{3.5cm} + \tfrac{(\frac{D}2 - 4) \Gamma(\frac{D}2)}{128 
\pi^{\frac{D}2}} \tfrac{H^4 a a' \Delta x^2}{[a a' \Delta x^2]^{\frac{D}2 - 2}} 
- \tfrac{k}{D (D-2)} H^2 a a' \Delta x^2 + \dots \qquad \label{Cdef} 
\end{eqnarray}
The second thing which makes this gauge so simple is that only a finite number
of terms survive for all three propagators in $D=4$ dimensions. We have already
seen this for the $A$ propagator (\ref{4Dtail}). The $D=4$ limits of the $B$ 
and $C$ propagators are even simpler,
\begin{equation}
D=4 \qquad \Longrightarrow \qquad i\Delta_B(x;x') = \tfrac1{4\pi^2} 
\tfrac{1}{a a' \Delta x^2} = i\Delta_C(x;x') \; . \label{4DBC}
\end{equation}

\subsection{Catalog of Large Loop Effects}

Ten graviton loops have so far been computed on de Sitter background
\cite{Tsamis:1996qm,Tsamis:1996qk,Tsamis:2005je,Miao:2005am,Kahya:2007bc,
Miao:2012bj,Leonard:2013xsa,Glavan:2015ura,Miao:2017vly,Glavan:2020gal,
Glavan:2021adm}. Of these, six show large logarithms in potentials and 
field strengths. In each case the Lagrangian is given, along with the 
appropriate linearized effective field equation and the enhanced field 
strength or exchange potential. For corrections to matter fields the 
matter loop corrections to gravity are also given. Theories are treated 
in order of increasing spin.

\subsubsection{Einstein-MMC Scalar}

The Lagrangian for a massless, minimally coupled (MMC) scalar is,
\begin{equation}
\mathcal{L}_{\rm MMC} = -\tfrac12 \partial_{\mu} \phi \partial_{\nu} \phi
g^{\mu\nu} \sqrt{-g} \; . \label{MMCS}
\end{equation}
The scalar's 1PI 2-point function is known as the self-mass 
$-i M_{\rm \scriptscriptstyle MMC}^2(x;x')$, and the 1-loop BPHZ counterterms 
required to subtract its divergences take the form, 
\begin{equation}
\Delta \mathcal{L}_{\rm MMC} = -\tfrac{\alpha_1}{2} \square \phi \square 
\phi \sqrt{-g} - \tfrac{\alpha_2}{2} R \partial_{\mu} \phi \partial_{\nu} \phi 
g^{\mu\nu} \sqrt{-g} - \tfrac{\alpha_3}{2} R \partial_{0} \phi \partial_{0}
\phi g^{00} \sqrt{-g} \; . \label{DMMCS}
\end{equation}
The original computation of $-i M_{\rm \scriptscriptstyle MMC}^2(x;x')$ 
\cite{Kahya:2007bc} was slightly corrected in \cite{Glavan:2021adm}.

The self-mass allows one to quantum-correct the linearized effective field
equation for the scalar,
\begin{equation}
\partial_{\mu} \Bigl[\sqrt{-g} \, g^{\mu\nu} \partial_{\nu} \phi(x)\Bigr] -
\int \!\! d^4x' \, M_{\rm \scriptscriptstyle MMC}^2(x;x') \phi(x') = J(x) 
\; . \label{ScalarEQN}
\end{equation}
This equation is nonlocal, but is real and causal ($M_{\rm \scriptscriptstyle 
MMC}^2(x;x') = 0$ for ${x'}^{\mu}$ outside the past light-cone of $x^{\mu}$) 
provided the Schwinger-Keldysh formalism is employed \cite{Chou:1984es,
Jordan:1986ug,Calzetta:1986ey,Ford:2004wc}. With the source $J(x)$ set to 
zero one is studying the propagation of dynamical scalars. Without loss of 
generality these can be taken to be spatial plane waves whose tree order mode 
function is, 
\begin{equation}
u^{\rm tree}(t,k) = \tfrac{H}{\sqrt{2 k^3}} [1 - \tfrac{i k}{a H}] 
\exp[\tfrac{i k}{a H}] = \tfrac{H}{\sqrt{2 k^3}} [1 + \tfrac1{2 a^2 H^2} +
\dots] \; . \label{utree}
\end{equation}
Note that it approaches a constant at late times. A loop of inflationary 
gravitons makes the approach to this constant slightly slower
\cite{Glavan:2021adm},
\begin{equation}
u(t,k) \longrightarrow u^{\rm tree} \Bigl\{ 1 + \tfrac{\kappa^2 k^2}{4 \pi^2} 
\tfrac{\ln[a]}{a^2} + \dots \Bigr\} \; . \label{scalarmode}
\end{equation}

It is also interesting to consider the response to a point source 
\cite{Glavan:2019yfc},
\begin{equation}
J(x) = K a \delta^3(\vec{x}) \qquad , \qquad \phi^{\rm tree} = \tfrac{HK}{4\pi}
\Bigl[-\tfrac1{a H r} + \ln(H r + \tfrac1{a})\Bigr] \; . \label{scalarsource}
\end{equation}
The tree order solution also approaches a constant at late times 
\cite{Glavan:2019yfc},
\begin{equation}
\phi^{\rm tree} = \tfrac{HK}{4\pi} \Bigl[\ln(H r) - \tfrac1{2 a^2 H^2 r^2}
+ \dots \Bigr] \longrightarrow \tfrac{HK}{4\pi} \ln(Hr) \; . \label{latepot}
\end{equation} 
The effect of a loop of inflationary gravitons is to add another logarithm at 
late times \cite{Glavan:2021adm},
\begin{equation}
\phi(x) \longrightarrow \tfrac{H K}{4\pi} \ln(H r) \Bigl\{1 - 
\tfrac{\kappa^2 H^2}{8 \pi^2} \ln(H r) + \dots \Bigr\} \; . 
\label{scalarpotential}
\end{equation}

The 1PI 2-point function for the graviton is known as the graviton 
self-energy $-i [\mbox{}^{\mu\nu} \Sigma^{\rho\sigma}](x;x')$. The BPHZ
counterterms needed to renormalize it at 1-loop are \cite{tHooft:1974toh},
\begin{equation}
\Delta \mathcal{L}_{\rm Einstein} = c_1 R^2 \sqrt{-g} + c_2 
C^{\alpha\beta\gamma\delta} C_{\alpha\beta\gamma\delta} \sqrt{-g} \; .
\label{DeltaEinstein}
\end{equation}
The single scalar loop contribution to the graviton self-energy was first
computed in \cite{Park:2011ww} and corrected slightly in \cite{Miao:2024atw}.

The graviton self-energy can be used to quantum-correct the linearized
Einstein equation,
\begin{equation}
\mathcal{L}^{\mu\nu\rho\sigma} \kappa h_{\rho\sigma}(x) - \int \!\! d^4x'
\, [\mbox{}^{\mu\nu} \Sigma^{\rho\sigma}](x;x') \kappa h_{\rho\sigma}(x')
= \tfrac{\kappa^2}{2} T^{\mu\nu}(x) \; , \label{EinsteinEQN}
\end{equation}
where the Lichnerowicz operator on de Sitter is,
\begin{eqnarray}
\lefteqn{\mathcal{L}^{\mu\nu\rho\sigma} h_{\rho\sigma} = \tfrac12 a^2 
\Bigl[\partial^2 h^{\mu\nu} \!-\! \eta^{\mu\nu} \partial^2 h \!+\! 
\eta^{\mu\nu} \partial^{\rho} \partial^{\sigma} h_{\rho\sigma} \!+\! 
\partial^{\mu} \partial^{\nu} h \!-\! 2 \partial^{\rho} 
\partial^{(\mu} h^{\nu)}_{~~\rho} \Bigr] } \nonumber \\
& & \hspace{1cm} + a^3 H \Bigl[ \eta^{\mu\nu} \partial_0 h \!-\! \partial_0
h^{\mu\nu} \!-\! 2 \eta^{\mu\nu} \partial^{\rho} h_{\rho 0} \!+\! 2 
\partial^{(\mu} h^{\nu)}_{~~0} \Bigr] \!+\! 3 a^4 H^2 \eta^{\mu\nu} h_{00} 
\; . \qquad \label{Lichnerowicz}
\end{eqnarray}
With the stress tensor $T^{\mu\nu}$ set to zero one can study plane wave
gravitons,
\begin{equation}
\kappa h_{\mu\nu} = \epsilon_{\mu\nu} e^{i \vec{k} \cdot \vec{x}} u(t,k) 
\; , \label{gravrad}
\end{equation}
where transverse-traceless and purely spatial polarization tensor 
$\epsilon_{\mu\nu}$ is the same as on flat space background and the tree 
order mode function is the same as (\ref{utree}). As with its scalar
cousin (\ref{scalarmode}), the scalar loop correction to the graviton
mode function falls off at late times, but it does enhance the electric
components of the Weyl tensor \cite{Miao:2024atw},
\begin{equation}
C_{0i0j} = C^{\rm tree}_{0i0j} \Bigl\{1 - \tfrac{3 \kappa^2 H^2}{160 \pi^2} 
\ln[a] + \dots\Bigr\} \; . \label{MMCWeyl}
\end{equation}

The gravitational response to a point mass requires two potentials,
\begin{equation}
T^{\mu\nu}(x) = -\delta^{\mu}_{~0} \delta^{\nu}_{~0} M a \delta^3(\vec{x}) 
\qquad , \qquad 
ds^2 = -[1 \!-\! 2 \Psi] dt^2 + a^2 [1 \!-\! 2 \Phi] d\vec{x} \!\cdot\!
d\vec{x} \; . \label{PointMass}
\end{equation}
The Newtonian potential is \cite{Miao:2024atw},
\begin{equation}
\Psi = \tfrac{G M}{a r} \Bigl\{1 + \tfrac{\kappa^2}{320 \pi^2 a^2 r^2} - 
\tfrac{3 \kappa^2 H^2}{160 \pi^2} \ln[a H r] + \dots\Bigr\} \; . 
\label{MMCNewton}
\end{equation}
The fractional correction proportional to $\kappa^2/a^2 r^2$ is the de
Sitter descendant of a well known flat space effect \cite{Radkowski:1970},
however, the secular $\kappa^2 H^2$ is new. Also new is the gravitational
slip \cite{Miao:2024atw},
\begin{equation}
\Psi + \Phi = \tfrac{G M}{a r} \Bigl\{0 + \tfrac{\kappa^2}{240 \pi^2 a^2 r^2} +
\tfrac{3 \kappa^2 H^2}{160 \pi^2} + \dots \Bigr\} \; . \label{MMCSlip} 
\end{equation}

Note from (\ref{MMCWeyl}) and (\ref{MMCNewton}) that a loop of inflationary
scalars weakens gravity. The physical interpretation for this seems to find
its origin in the vast ensemble of scalars which are produced by inflation.
The gravitational sector is weakened because it must supply the energy to
produce these particles.

\subsubsection{Einstein-MCC Scalar}

The Lagrangian for a massless, conformally coupled (MCC) scalar is,
\begin{equation}
\mathcal{L}_{\rm MCC} = -\tfrac12 \partial_{\mu} \phi \partial_{\nu} \phi
g^{\mu\nu} \sqrt{-g} - \tfrac18 (\tfrac{D-2}{D-1}) \phi^2 R \sqrt{-g} \; .
\label{MCCS}
\end{equation}
The MCC self-mass $-i M^2_{\rm \scriptscriptstyle MCC}(x;x')$ is the 
scalar's 1PI 2-point function. It was first computed in \cite{Boran:2014xpa,
Boran:2017fsx}, however, a problem was noted in the flat space limit 
\cite{Frob:2017apy} which was corrected in \cite{Glavan:2020gal}.

As the name suggests, $\mathcal{L}_{\rm MCC}$ is invariant under a 
conformal (Weyl) transformation in any dimension $D$,
\begin{equation}
\phi(x) \longrightarrow \Omega^{\frac{D}2 -1}(x) \times \phi(x) \qquad ,
\qquad g_{\mu\nu}(x) \longrightarrow \Omega^2(x) \times g_{\mu\nu}(x) \; .
\label{WeylTrans}
\end{equation}
This means that taking the conformal factor $\Omega$ equal to the 
inverse scale factor $1/a(t)$, and working in conformal coordinates,
erases any dependence on the scale factor when the conformally 
transformed fields $\widetilde{\phi}$ and $\widetilde{g}_{\mu\nu}$ 
are used,
\begin{equation}
\mathcal{L}_{\rm MCC} = -\tfrac12 \partial_{\mu} \widetilde{\phi} 
\partial_{\nu} \widetilde{\phi} \, \widetilde{g}^{\mu\nu} 
\sqrt{-\widetilde{g}} - \tfrac18 (\tfrac{D-2}{D-1}) {\widetilde{\phi}}^2 
\widetilde{R} \sqrt{-\widetilde{g}} \; . \label{ConformalFields}
\end{equation}
Note that, for any cosmological geometry (which implies $\widetilde{R} = 0$), 
this Lagrangian is invariant under constant shifts $\widetilde{\phi}(x) 
\rightarrow \widetilde{\phi}(x) + C$. If we insist that counterterms are 
invariant under this same shift symmetry then renormalization of 
$-i M^2_{\rm \scriptscriptstyle MMC}(x;x')$ requires three terms 
\cite{Glavan:2020gal},
\begin{eqnarray}
\lefteqn{ \Delta \mathcal{L}_{\rm MMC} = -\tfrac{\alpha}{2} \Bigl[ \square
\phi - \tfrac14 (\tfrac{D-2}{D-1}) R \phi \Bigr]^2 \sqrt{-g} - \tfrac{\beta}{2}
\Bigl[ \square \phi - \tfrac14 (\tfrac{D-2}{D-1}) R \phi]\Bigr] \tfrac{\phi R
\sqrt{-g}}{D (D-1)} } \nonumber \\
& & \hspace{9cm} -\tfrac{\gamma}{2} \partial_i \phi \partial_j \phi 
\tfrac{R \sqrt{-g}}{D (D-1)} , \qquad \label{DeltaMCC} \\
& & \hspace{0.5cm} \longrightarrow -\tfrac{\alpha}{2 a^2} (\partial^2 
\widetilde{\phi})^2 + \tfrac{\beta H^2}{2} \partial_{\mu} \widetilde{\phi} \,
\partial^{\mu} \widetilde{\phi} - \tfrac{\gamma H^2}{2} \partial_i
\widetilde{\phi} \, \partial_i \widetilde{\phi} \qquad {\rm (de\ Sitter)} \; . 
\qquad \label{DeltaMCCconf}  
\end{eqnarray}
The noncovariant term proportional to $\gamma$ derives from the de Sitter 
breaking gauge.

The scalar self-mass can be used to quantum-correct the linearized effective
field equation for the scalar,
\begin{equation}
\partial_{\mu} \Bigl[\sqrt{-g} \, g^{\mu\nu} \partial_{\nu} \phi(x)\Bigr] -
\tfrac1{6} \sqrt{-g} \, R \phi(x) - \int \!\! d^4x' \, M_{\rm 
\scriptscriptstyle MCC}^2(x;x') \phi(x') = J(x) \; . \label{MCCSEQN}
\end{equation}
When this is solved (with $J(x) = 0$) for spatial plane wave radiation the
result at late times is a secular enhancement of the mode function 
\cite{Glavan:2020ccz}, 
\begin{equation}
u(t,k) \longrightarrow \tfrac1{a \sqrt{2 k}} \exp[\tfrac{i k}{a H}] \Bigl\{1
+ \tfrac{\kappa^2 H^2}{24 \pi^2} \ln(a) + \dots \Bigr\} \; . \label{MCCmode}
\end{equation}
Note that the tree order conformally coupled scalar mode function redshifts
to zero, quite unlike its minimally coupled cousin (\ref{utree}). When
equation (\ref{MCCSEQN}) is solved with $J(x) = K a \delta^3(\vec{x})$ the
result is a similar late time enhancement \cite{Glavan:2020ccz}, 
\begin{equation}
\phi(x) \longrightarrow -\tfrac{K}{4\pi a r} \Bigl\{1 + 
\tfrac{\kappa^2}{48 \pi^2 a^2 r^2} + \tfrac{\kappa^2 H^2}{24 \pi^2} 
\ln(a H r) + \dots \Bigr\} \; . \label{MCCphi}
\end{equation}
It should be noted that both enhancements (\ref{MCCmode}) and (\ref{MCCphi})
could be absorbed into a finite counterterm of the form $\phi^2 R^2 
\sqrt{-g}$ if one does not require (\ref{DeltaMCC}) to respect shift 
symmetry for de Sitter.

The conformal scalar loop correction to the graviton self-energy is very
simple to compute. The fact that $\mathcal{L}_{\rm MCC}$ has no dependence 
on the scale factor when expressed in terms of conformal fields 
(\ref{ConformalFields}) means that the primitive result is identical to
the flat space computation which was made half a century ago 
\cite{Capper:1973bk}. The only dependence on the scale factor comes from
renormalization \cite{Foraci:2024cwi}. When the resulting self-energy is used
to quantum-correct the linearized Einstein equation (\ref{EinsteinEQN}) on
de Sitter background the 1-loop corrections to plane wave radiation and 
the response to a point mass are,
\begin{eqnarray}
C_{0i0j} &\!\!\! = \!\!\!& C^{\rm tree}_{0i0j} \Bigl\{1 + 
\tfrac{\kappa^2 H^2}{480 \pi^2} \ln(a) + \dots\Bigr\} \; , \qquad 
\label{MCCWeyl} \\
\Psi &\!\!\! = \!\!\!& \tfrac{G M}{a r} \Bigl\{1 + 
\tfrac{\kappa^2}{720 \pi^2 a^2 r^2} + \tfrac{\kappa^2 H^2}{480 \pi^2} 
\ln(a H r) + \dots\Bigr\} \; , \qquad \label{MCCNewton} \\
\Psi + \Phi &\!\!\! = \!\!\!& \tfrac{G M}{a r} \Bigl\{0 + 
\tfrac{\kappa^2}{1440 \pi^2 a^2 r^2} - \tfrac{\kappa^2 H^2}{480 \pi^2} 
+ \dots \Bigr\} \; . \qquad \label{MCCSlip} 
\end{eqnarray}

Unlike the minimally coupled scalar, a loop of conformally coupled scalars
strengthens gravity. The exact conformal invariance of $\mathcal{L}_{\rm 
MCC}$ means that none of these scalars are created during inflation. Their
effect on gravity seems rather to come from the redshift of their vacuum
energies which goes to enhance the gravitational sector.

\subsubsection{Einstein-Dirac}

The Dirac Lagrangian is (\ref{Einstein}) plus,
\begin{equation}
\mathcal{L}_{\rm Dirac} = \overline{\psi} e^{\mu}_{~a} \gamma^a \Bigl(i 
\partial_{\mu} \!-\! \tfrac12 A_{\mu bc} J^{bc} \Bigr) \psi \sqrt{-g} \; . 
\label{Dirac}
\end{equation}
Here $\gamma^a$ are the gamma matrices, $e^{\mu}_{~a}(x)$ is the vierbein 
field, $A_{\mu bc}(x)$ is the spin connection and $J^{bc}$,
\begin{equation}
g^{\mu\nu} = e^{\mu}_{~a} e^{\nu}_{~b} \eta^{ab} \quad , \quad 
A_{\mu bc} \equiv e^{\nu}_{~b} (e_{\nu c , \mu} - \Gamma^{\rho}_{~\mu\nu}
e_{\rho c}) \quad , \quad J^{bc} \equiv \tfrac{i}{4} [\gamma^{b}, 
\gamma^{c}] \; . \label{spinstuff}
\end{equation}
If the local Lorentz gauge is fixed by requiring the vierbein to be 
symmetric there are no local Lorentz ghosts and one can regard the vierbein 
as a function of the graviton field \cite{Woodard:1984sj},
\begin{equation}
e_{\mu b} = a \Bigl[\sqrt{\widetilde{g} \eta^{-1}} \,\Bigr]_{\mu}^{~c} 
\eta_{c b} = a \Bigl[\eta_{\mu b} + \tfrac{\kappa}{2} h_{\mu b} - 
\tfrac{\kappa^2}{8} h_{\mu}^{~ c} h_{c b} + \tfrac{\kappa^3}{16} 
h_{\mu}^{~c} h_{c}^{~d} h_{d b} - \dots\Bigr] . \label{vierbein}
\end{equation}

Because $\mathcal{L}_{\rm Dirac}$ is conformally invariant for any
dimension $D$, it has no dependence on the scale factor at all on
the scale factor when both the fermion and the metric are conformally
rescaled, 
\begin{equation}
\psi \equiv a^{\frac{D-1}{2}} \widetilde{\psi} \;\; , \;\; g_{\mu\nu}
\equiv a^2 \widetilde{g}_{\mu\nu} \;\; , \;\; \mathcal{L}_{\rm Dirac} = 
\overline{\widetilde{\psi}} \widetilde{e}^{\mu}_{~a} \gamma^a \Bigl(i 
\partial_{\mu} \!-\! \tfrac12 \widetilde{A}_{\mu bc} J^{bc} \Bigr) 
\widetilde{\psi} \sqrt{-\widetilde{g}} \; . \label{newDirac}
\end{equation}
This means that dependence on the scale factor can only enter through
the graviton propagator and through counterterms. The (conformally
transformed) 1PI 2-point function for $\widetilde{\psi}$ is known as
the fermion self-energy $-i [\mbox{}_{i} S_{j}](x;x')$ and the graviton
loop contribution to it on de Sitter background was first computed in
\cite{Miao:2005am}. To express the BPHZ counterterms by which it is 
renormalized we define the spin covariant derivative as,
\begin{equation}
\mathcal{D}_{\mu} \equiv \partial_{\mu} + \tfrac{i}{2} A_{\mu bc} J^{bc}
\; . \label{Dspin}
\end{equation}
In these terms the resulting counterterms are \cite{Miao:2005am},
\begin{eqnarray}
\lefteqn{\Delta \mathcal{L}_{\rm Dirac} = \alpha_1 \overline{\psi} 
(e^{\mu}_{~a} \gamma^a i \mathcal{D}_{\mu})^3 \psi \sqrt{-g} + \alpha_2 R 
\overline{\psi} e^{\mu}_{~a} \gamma^a i \mathcal{D}_{\mu} \psi \sqrt{-g} }
\nonumber \\
& & \hspace{7.8cm} + \alpha_3 R \overline{\psi} e^{k}_{~\ell} \gamma^{\ell} 
i \mathcal{D}_{k} \psi \sqrt{-g} \; . \qquad \label{DeltaDirac}
\end{eqnarray}
If we specialize the de Sitter background and express the counterterms
using the conformally transformed field this becomes,
\begin{eqnarray}
\lefteqn{\Delta \mathcal{L}_{\rm Dirac} \longrightarrow \alpha_1 
\overline{\widetilde{\psi}} (i \gamma^{\mu} \partial_{\mu} a^{-1})^3 a 
\widetilde{\psi} + \alpha_2 D (D\!-\!1) H^2 \overline{\widetilde{\psi}} 
i \gamma^{\mu} \partial_{\mu} \widetilde{\psi} } \nonumber \\
& & \hspace{7.5 cm} + \alpha_3 D (D \!-\! 1) H^2 \overline{\widetilde{\psi}} 
i \gamma^{k} \partial_{k} \widetilde{\psi} \; . \qquad \label{DeltaDiracdS}
\end{eqnarray}

The fermion self-energy can be used to quantum-correct the conformally
transformed Dirac equation,
\begin{equation}
i \gamma^{\mu}_{ij} \partial_{\mu} \widetilde{\psi}_{j}(x) - \int \!\!
d^4x' \, [\mbox{}_i S_j](x;x') \widetilde{\psi}_{j}(x') = 0 \; . 
\label{DiracEQN}
\end{equation}
When this is solved for plane wave fermions the 1-loop correction induces,
a secular enhancement of the field strength \cite{Miao:2005am},
\begin{equation}
\widetilde{\psi}(x) = \widetilde{\psi}^{\rm tree} \Bigl\{1 + 
\tfrac{17 \kappa^2 H^2}{128 \pi^2} \ln[a] + \dots \Bigr\} \; . 
\label{GRDirac}
\end{equation}
Although it would simple enough to include a source on the right hand 
side of equation (\ref{DiracEQN}), the response to it does not seem to
give the potential for exchanging a fermion. The long range force 
induced by massless fermions requires the exchange of {\it two} fermions 
\cite{Feinberg:1968zz,Hsu:1992tg} and is a 2-loop phenomenon beyond the 
range of current computation.

The single fermion loop correction to the graviton self-energy turns out
to be $6$ times the conformal scalar one \cite{Capper:1973mv,Duff:2000mt,
Foraci:2024cwi}. This means that one can use the conformal scalar results 
(\ref{MCCWeyl}-\ref{MMCSlip}) to obtain those for a Dirac fermion,
\begin{eqnarray}
C_{0i0j} &\!\!\! = \!\!\!& C_{0i0j}^{\rm tree} \Bigl\{ 1 +
\tfrac{\kappa^2 H^2}{80 \pi^2} \ln[a] + \dots \Bigr\} \; , \qquad 
\label{DiracWeyl} \\
\Psi &\!\!\! = \!\!\!& \tfrac{G M}{a r} \Bigl\{1 + \tfrac{\kappa^2}{120 \pi^2
a^2 r^2} - \tfrac{\kappa^2 H^2}{80 \pi^2} \ln[a H r] + \dots \Bigr\} \; ,
\qquad \label{DiracNewton} \\
\Psi + \Phi &\!\!\! = \!\!\!& \tfrac{G M}{a r} \Bigl\{0 - \tfrac{\kappa^2}{
240 \pi^2 a^2 r^2} - \tfrac{\kappa^2 H^2}{80 \pi^2} + \dots \Bigr\} \; . 
\qquad \label{DiracSlip}
\end{eqnarray}
Like the conformal scalar case, the effect of a loop of fermions is to 
strengthen gravity during inflation. This might seem surprising because
the exact conformal invariance of $\mathcal{L}_{\rm Dirac}$ means that
no fermions are created during inflation and fermion vacuum 0-point 
energies are negative. However, it was long ago shown that another minus
sign enters the coupling to gravity \cite{Capper:1973mv}, so conformal
scalars and massless fermions both strengthen the gravitational sector.

\subsubsection{Einstein-Maxwell}

The Lagrangian for electromagnetism is,
\begin{equation}
\mathcal{L}_{\rm Maxwell} = -\tfrac14 F_{\rho\mu} F_{\sigma\nu} 
g^{\rho\sigma} g^{\mu\nu} \sqrt{-g} \label{Maxwell}
\end{equation}
The 1PI 2-point function for the photon is known as the vacuum
polarization $i [\mbox{}^{\mu} \Pi^{\nu}](x;x')$. The single graviton
contribution to it was first computed on de Sitter background in 
\cite{Leonard:2013xsa}. Although Einstein-Maxwell is not renormalizable
\cite{Deser:1974cz,Deser:1974zzd}, the BPHZ counterterms needed to 
renormalize $i [\mbox{}^{\mu} \Pi^{\nu}](x;x')$ are \cite{Leonard:2013xsa},
\begin{eqnarray}
\lefteqn{\mathcal{L}_{\rm Maxwell} = C_4 D_{\alpha} F_{\mu\nu} D_{\beta} 
F_{\rho\sigma} g^{\alpha\beta} g^{\mu\rho} g^{\nu\sigma} \sqrt{-g}
+ \overline{C} H^2 F_{\mu\nu} F_{\rho\sigma} g^{\mu\rho} g^{\nu\sigma}
\sqrt{-g} } \nonumber \\
& & \hspace{7.4cm} + \Delta C H^2 F_{ij} F_{k\ell} g^{ik} g^{j\ell}
\sqrt{-g} \; . \qquad \label{DeltaMaxwell}
\end{eqnarray} 

The vacuum polarization can be used to quantum-correct Maxwell's
equation,
\begin{equation}
\partial_{\nu} \Bigl[ \sqrt{-g} \, g^{\nu\rho} g^{\mu\sigma} 
F_{\rho\sigma}(x) \Bigr] + \int \!\! d^4x' \, [\mbox{}^{\mu} 
\Pi^{\rho\sigma}](x;x') A_{\nu}(x') = J^{\mu}(x) \; . \label{PhotonEQN}
\end{equation}
With the current density set to zero the solutions correspond to
radiation. The result for the electric field strength of a plane 
wave photon shows a secular enhancement \cite{Wang:2014tza},
\begin{equation}
F_{0i} = F_{0i}^{\rm tree} \Bigl\{1 + \tfrac{\kappa^2 H^2}{8 \pi^2} \ln[a] 
+ \dots \Bigr\} \; . \label{GRElectric}
\end{equation}

It is also interesting to work out the Coulomb potential, which is the
response to a static point charge,
\begin{equation}
J^{\mu}(x) = \delta^{\mu}_{~0} Q \delta^3(\vec{x}) \; . \label{PointCharge}
\end{equation}
The single graviton loop correction to the Coulomb potential is
\cite{Glavan:2013jca},
\begin{equation}
\Phi = \tfrac{Q}{4\pi a r} \Bigl\{1 + \tfrac{\kappa^2}{24 \pi^2 a^2 r^2} + 
\tfrac{\kappa^2 H^2}{8 \pi^2} \ln[a H r] + \dots \Bigr\} \; . 
\label{GRCoulomb}
\end{equation}
The fractional correction proportional to $\kappa^2/a^2 r^2$ is the de 
Sitter version of a flat space effect first derived in 1970 
\cite{Radkowski:1970}. The secular effect proportional to $\kappa^2 H^2$
arises from inflationary gravitons.

The single photon contribution to the graviton self-energy on de Sitter
was first computed on flat space in \cite{Capper:1974ed}, and on de Sitter
in \cite{Wang:2015eaa}. It can be renormalized by the same BPHZ counterterms 
(\ref{DeltaEinstein}) as for any matter loop. When the graviton self-energy 
is used to quantum-correct the linearized Einstein equation (\ref{EinsteinEQN}), 
the results for the electric components to the Weyl tensor \cite{Foraci:2024vng}
and the Newtonian potential \cite{Wang:2015eaa} are,
\begin{eqnarray}
C_{0i0j} &\!\!\! = \!\!\!& C_{0i0j}^{\rm tree} \Bigl\{ 1 + 
\tfrac{\kappa^2 H^2}{40 \pi^2} \ln[a] + \dots \Bigr\} \; , \qquad 
\label{EMWeyl} \\
\Psi &\!\!\! = \!\!\!& \tfrac{G M}{a r} \Bigl\{1 + \tfrac{\kappa^2}{60 \pi^2
a^2 r^2} + \tfrac{\kappa^2 H^2}{40 \pi^2} \ln[a H r] + \dots \Bigr\} \; .
\qquad \label{EMNewton}
\end{eqnarray}
The gravitational slip is \cite{Wang:2015eaa},
\begin{equation}
\Psi + \Phi = \tfrac{G M}{a r} \Bigl\{0 + \tfrac{\kappa^2}{120 \pi^2 a^2 r^2} +
\tfrac{\kappa^2 H^2}{40 \pi^2} + \dots \Bigr\} \; . \label{EMSlip} 
\end{equation}

From expressions (\ref{EMWeyl}) and (\ref{EMNewton}) we see that a loop of
photons strengthens gravity during inflation. Because electromagnetism is
conformally invariant in $D=4$ dimensions there is no production of photons
during inflation. The sign of the 1-loop effect seems to derive from the
same redshift of the photon 0-point energy which explains the conformal scalar
and massless fermion results.

\subsubsection{Pure Einstein}

The graviton self-energy $-i [\mbox{}^{\mu\nu} \Sigma^{\rho\sigma}](x;x')$ 
was computed in $D=4$ for ${x'}^{\mu} \neq x^{\mu}$ in \cite{Tsamis:1996qk}.
Because dimensional regularization was not employed, the list of required 
counterterms is not yet known although it must include (\ref{DeltaEinstein}),
along with certain noninvariants (such as the final terms of (\ref{DeltaL}),
(\ref{DeltaDirac}) and (\ref{DeltaMaxwell})) arising from the de Sitter 
breaking gauge. Although the graviton self-energy obeys the Ward identity 
away from coincidence \cite{Tsamis:1996qk}, the existence of a delta function 
obstacle for the scalar contribution \cite{Tsamis:2023fri}, suggests that full 
conservation will probably require a finite renormalization of the cosmological 
constant.

An attempt was made to reconstruct the renormalized graviton self-energy 
from the $D=4$ result \cite{Tan:2021ibs}. Because the calculation was done
in a fixed gauge, the classical parts of the linearized Einstein equation
must be written in this same gauge,
\begin{equation}
\mathcal{D}^{\mu\nu\rho\sigma} \kappa h_{\rho\sigma}(x) - \int \!\! d^4x' \,
[\mbox{}^{\mu\nu} \Sigma^{\rho\sigma}](x;x') \kappa h_{\rho\sigma}(x') = 
\tfrac{\kappa^2}{2} T^{\mu\nu}(x) \; . \label{EinsteinGaugeEQN}
\end{equation} 
Here the kinetic operator in the gauge (\ref{LGF}) is,
\begin{equation}
\mathcal{D}^{\mu\nu\rho\sigma} = [\tfrac12 \eta^{\mu (\rho} \eta^{\sigma) \nu}
- \tfrac14 \eta^{\mu\nu} \eta^{\rho\sigma}] \partial^{\mu} a^2 \partial_{\mu}
+ 2 a^2 H^2 \delta^{(\mu}_{~~0} \eta^{\nu) (\rho} \delta^{\sigma)}_{~~0} \; .
\label{GaugeKinetic}
\end{equation}

When equation (\ref{EinsteinGaugeEQN}) is solved for gravitational radiation
the solution takes the form (\ref{gravrad}) with a mode function which is
strongly enhanced at late times \cite{Tan:2021lza},
\begin{equation}
u(t,k) \longrightarrow u^{\rm tree} \Bigl\{1 + \tfrac{\kappa^2 H^2}{3 \pi^2}
\ln^2(a) + \dots \Bigr\} \; . \label{GRmode}
\end{equation}
The Newtonian potential is even more strongly enhanced \cite{Tan:2022xpn},
\begin{equation}
\Psi = \frac{G M}{a r} \Bigl\{1 + \tfrac{103 \kappa^2}{240 \pi^2 a^2 r^2} 
+ \tfrac{\kappa^2 H^2}{2 \pi^2} \Bigl[\ln^3(a) - 3 \ln(a) \ln(H r) \Bigr] 
+ \dots \Bigr\} \; . \label{GRNewton}
\end{equation}
And the gravitational slip is also enhanced \cite{Tan:2022xpn},
\begin{equation}
\Psi + \Phi = \frac{G M}{a r} \Bigl\{0 + \tfrac{7 \kappa^2}{20 \pi^2 a^2 r^2} 
- \tfrac{2 \kappa^2 H^2}{\pi^2} \Bigl[\ln^2(a) - \tfrac34 \ln(H r) \Bigr] 
+ \dots \Bigr\} \; . \label{GRSlip}
\end{equation}

These results require confirmation by a dimensionally regulated and fully
renormalized re-computation of the graviton self-energy, with proper 
consideration of possible need for the same sort of finite renormalization
of the cosmological constant which scalars require \cite{Tsamis:2023fri}.
If correct, both the mode function (\ref{GRmode}) and the Newtonian 
potential (\ref{GRNewton}) show an unprecedented strengthening. One possible
reason for this is that the $\kappa h \partial h \partial h$ interaction of
gravity allows a loop of two gravitons to involve more tail terms 
(\ref{4Dtail}) than any of the other cases so far considered. 

\section{Modified Stochastic Formalism}

The point of this section is to show how Starobinsky's stochastic 
formalism \cite{Starobinsky:1986fx,Starobinsky:1994bd} can be adapted
to describe and resum the large loop corrections engendered by the tail
part (\ref{4Dtail}) of the graviton propagator. The section begins by
reviewing the case of scalar potential models for which the formalism
was originally developed. Then consideration moves to other fields
whose propagators lack a tail. Derivative interactions such as the 
fundamental $\kappa h \partial h \partial h$ vertex of gravity are
approached gradually, first by nonlinear sigma models and then by 
matter fields. The section concludes with a discussion of gravity 
itself.

\subsection{\normalsize Scalar Potential Models}

Fields with the logarithmic tail (\ref{4Dtail}) are known as ``active'';
other fields are ``passive''. The simplest active field is a scalar 
potential model,
\begin{equation}
\mathcal{L} = -\tfrac12 \partial_{\mu} \phi \partial_{\nu} \phi 
g^{\mu\nu} \sqrt{-g} - V(\phi) \sqrt{-g} \; . \label{scalpot}
\end{equation}
Its leading logarithms can be captured by replacing the Heisenberg 
equation for the (dimensionally regulated) full field $\phi(x)$, 
with the Langevin equation for the UV finite (hence $D=4$) stochastic 
random variable $\varphi(x)$ \cite{Starobinsky:1986fx},
\begin{equation}
\ddot{\phi} + (D\!-\!1) H \dot{\phi} - \tfrac{\nabla^2}{a^2} \phi = -V'(\phi) 
\qquad \longrightarrow \qquad 3 H [ \dot{\varphi} - \dot{\varphi}_0 ] = 
-V'(\varphi) \; . \label{Langevin}
\end{equation}
The stochastic ``jitter'' $\dot{\varphi}_0(x)$ is the time 
derivative of the infrared-truncated free field,
\begin{equation}
\varphi_0(x) \equiv \int_{H}^{Ha} \!\!\!\! \tfrac{d^3k}{(2\pi)^3} 
\tfrac{H e^{i \vec{k} \cdot \vec{x}} }{\sqrt{2 k^3}} \, [ \alpha(\vec{k}) 
+ \alpha^{\dagger}(-\vec{k})] \;\; , \;\; [\alpha(\vec{k}), 
\alpha^{\dagger}(\vec{k}')] = (2\pi)^3 \delta^3(\vec{k} - \vec{k}') 
\; . \label{jitter}
\end{equation}
Because $\varphi_0(x)$ commutes with its time derivative it behaves as 
a stochastic random variable; because the mode sum is truncated at $k =
H a$, correlators of it are completely ultraviolet finite. The same two 
properties are inherited by $\varphi(x)$. 

The fact that (ultraviolet divergent) correlators of the full field 
$\phi(x)$ agree, at leading logarithm order, with those of $\varphi(x)$ 
can be proved \cite{Tsamis:2005hd}. One first acts the inverse kinetic 
operator of the full field equation to the reach the Yang-Feldman 
equation, with free field $\phi_0(x)$,
\begin{equation}
\phi(x) = \phi_0(x) - \int \!\! d^4x' \sqrt{-g(x')} \, i \theta(t \!-\! t')
[\phi_0(x) , \phi_0(x')] V'\Bigl( \phi(x')\Bigr) \; . \label{YangFeldman}
\end{equation}
Equation (\ref{YangFeldman}) is still exact. Iterating it would give the 
interaction picture expansion of the full field $\phi$ in terms $\phi_0$. 
Because reaching leading logarithm order requires a logarithm be contributed
by each pair of free fields (including retarded propagators, which involve 
the commutator $[\phi_0(x),\phi_0(x')]$), there is no change, at leading 
logarithm order, by cutting off the free field mode sum at $k = H a(t)$ 
and truncating the mode function to its leading nonzero contribution for 
small $k$,
\begin{equation}
u(t,k) = \tfrac{H}{\sqrt{2 k^3}} [1 - \tfrac{i k}{Ha}] e^{i\frac{k}{Ha}}
\longrightarrow \tfrac{H}{\sqrt{2 k^3}} [1 + \tfrac{1}{2} (\tfrac{k}{Ha})^2
+ \tfrac{i}{3} (\tfrac{k}{Ha})^3 + \dots ] \; . \label{modefunc}
\end{equation}
This means retaining just the first term for the initial factor of 
$\phi_0(x)$ and one factor of the third term for the commutator. The
result is an equation for the stochastic random field $\varphi(x)$, whose
correlators agree with those of $\phi(x)$ at leading logarithm order,
\begin{equation}
\varphi(t,\vec{x}) = \varphi_0(t,\vec{x}) - \int \!\! d^4x' 
\tfrac{\theta(t - t')}{3 H} \delta^3(\vec{x} \!-\! \vec{x}') V'\Bigl(
\varphi(t,\vec{x})\Bigr) \; . \label{stochtrunc}
\end{equation}
Differentiation of (\ref{stochtrunc}) with respect to $t$ gives Starobinsky's
Langevin equation (\ref{Langevin}), which completes the proof. Note the
essential role played by the fact that reaching leading logarithm order
requires that each pair of free fields must contribute a large logarithm. 
This is what fails when passive fields or derivative interactions are present.

Passive fields such as fermions and photons do not induce large logarithms
but they can mediate logarithms inherited from active fields. Examples 
include a photon which is minimally coupled to a complex scalar and a 
fermion which is Yukawa-coupled to a real scalar,
\begin{eqnarray}
\lefteqn{\mathcal{L}_{\rm Yukawa} = \overline{\psi} e^{\mu}_{~a} \gamma^a 
\Bigl(i \partial_{\mu} - \tfrac12 A_{\mu bc} J^{bc} \Bigr) \psi \sqrt{-g} }
\nonumber \\
& & \hspace{3.5cm} -\tfrac12 \partial_{\mu} \phi \partial_{\nu} \phi g^{\mu\nu} 
\sqrt{-g} - V(\phi) \sqrt{-g} - f \phi \overline{\psi} \psi \sqrt{-g}  
\; . \qquad \label{Yukawa} 
\end{eqnarray}
Passive fields make nonzero contributions to correlators which derive 
as much from the ultraviolet as from the infrared, and from their full mode 
functions, so stochastically truncating them would represent a serious error. 
The correct procedure is to instead integrate out passive fields in the 
presence of a spacetime constant active field background \cite{Miao:2006pn,
Prokopec:2007ak}.

For example, the scalar equation for Yukawa theory (\ref{Yukawa}) is,
\begin{equation}
\partial_{\mu} (\sqrt{-g} g^{\mu\nu} \partial_{\nu} \phi) - V'(\phi) \sqrt{-g} 
- f \overline{\psi} \psi \sqrt{-g} = 0 \; . \label{YEOM}
\end{equation}
Integrating out the fermion field in the presence of constant $\phi$ 
gives the trace of the coincidence limit of the fermion propagator with
mass $m = f \phi$,
\begin{equation}
f \overline{\psi}(x) \psi(x) \longrightarrow -f [\mbox{}_{i} S_{i}](x;x)
\equiv V'_{\rm eff}(\phi) \; . \label{Veff}
\end{equation}
The massive fermion propagator is known on de Sitter background 
\cite{Candelas:1975du}, and the divergences of its coincidence limit can
be absorbed by quadratic and quartic counterterms in the potential
$V(\phi)$. After the divergences have been removed the result is a scalar
potential model which can be treated using Starobinsky's stochastic
formalism \cite{Starobinsky:1986fx,Starobinsky:1994bd}. The results agree
at leading logarithm order with exact computations in the original theory
\cite{Miao:2006pn}. The same procedure has been implemented for SQED
(Scalar Quantum Electrodynamics) \cite{Prokopec:2007ak}, using the massive 
photon propagator on de Sitter \cite{Tsamis:2006gj}. Like the case of 
Yukawa, the stochastic formulation of SQED has been shown to reproduce the 
leading logarithms of exact dimensionally regulated and fully renormalized 
results at 1-loop and 2-loop orders \cite{Prokopec:2006ue,Prokopec:2008gw}.

\subsection{\normalsize Nonlinear Sigma Models}

A simple nonlinear sigma model is \cite{Miao:2021gic},
\begin{equation}
\mathcal{L} = -\tfrac12 \partial_{\mu} A \partial_{\nu} A g^{\mu\nu} \sqrt{-g}
- \tfrac12 (1 \!+\! \tfrac{\lambda}2 A)^2 \partial_{\mu} B \partial_{\nu} B 
g^{\mu\nu} \sqrt{-g} \; . \label{LAB}
\end{equation}
The $A \partial B \partial B$ interaction in (\ref{LAB}) bears an 
obvious similarity to the fundamental $h \partial h \partial h$ interaction of
gravity. This analogy is why nonlinear sigma models have long been employed 
to study derivative interactions without the distractions of tensor indices 
and the gauge problem \cite{Tsamis:2005hd,Kitamoto:2010et,Kitamoto:2011yx,
Kitamoto:2018dek}.

Interactions involving derivatives of active fields introduce a new 
complication. Like passive fields, correlators which include them 
receive contributions from ultraviolet and from the full mode function. To 
see the problem, consider the coincidence limit of the doubly 
differentiated propagator \cite{Onemli:2002hr},
\begin{equation}
\lim_{x' \rightarrow x} \partial_{\mu} \partial'_{\nu} i\Delta(x;x') =
\langle \Omega \vert \partial_{\mu} \phi_0(x) \partial_{\nu} \phi_0(x) 
\vert \Omega \rangle = -\tfrac{H^D}{(4\pi)^\frac{D}2} \tfrac{\Gamma(D)}{
2 \Gamma(\frac{D}2 \!+\! 1)} \, g_{\mu\nu}(x) \; . \label{ddprop}
\end{equation}
Replacing the full free field $\phi_0(x)$ by its stochastic cognate 
$\varphi_0(x)$ in (\ref{ddprop}) would produces a positive, noncovariant 
result for the case of $\mu = \nu$ because it is the square a finite, 
Hermitian operator. In contrast, the exact, dimensionally regulated result 
(\ref{ddprop}) is covariant, and negative for spatial indices.

The correct procedure is to integrate differentiated active fields in the 
presence of a constant active field background. For (\ref{LAB}) it is the 
$B$ fields which are differentiated and they can be integrated out because 
setting $A$ to a spacetime constant just changes the $B$ field strength,
\begin{equation}
\langle \Omega \vert T[B(x) B(x')]\vert \Omega \rangle_{A_0} =
\tfrac{i \Delta(x;x')}{(1 + \frac{\lambda}{2} A_0)^2} \; . \label{fieldS}
\end{equation}
Hence one can eliminate the differentiated $B$ fields from the $A$ 
equation,
\begin{equation}
\partial_{\mu} ( \sqrt{-g} \, g^{\mu\nu} \partial_{\nu} A) = \tfrac{\lambda}{2} 
(1 \!+\! \tfrac{\lambda}2 A) \partial_{\mu} B \partial_{\nu} B g^{\mu\nu} 
\sqrt{-g} \longrightarrow -\tfrac{3 \lambda H^4}{16 \pi^2} \tfrac{\sqrt{-g}}{1 
\!+\! \frac{\lambda}2 A} \; . \label{Aeqn}
\end{equation}
This is a scalar potential model and Starobinsky's procedure 
\cite{Starobinsky:1986fx,Starobinsky:1994bd} can be invoked to find the 
Langevin equation for the associated stochastic random field 
$\mathcal{A}(t,\vec{x})$ whose leading logarithms agree with those of 
$A(t,\vec{x})$,
\begin{equation}
3 H ( \dot{\mathcal{A}} - \dot{\mathcal{A}}_0 ) = \tfrac{3 \lambda H^4}{16 \pi^2} 
\tfrac1{1 \!+\! \frac{\lambda}2 \mathcal{A}} \; . \label{ALangevin}
\end{equation}
Here $\mathcal{A}_0$ is the same infrared truncation as (\ref{jitter}).
The solution of (\ref{ALangevin}) consists of a $\comp$-number function of 
time plus a series of terms involving powers of the stochastic jitter 
$\mathcal{A}_0$,
\begin{eqnarray}
\lefteqn{ \mathcal{A}(t,\vec{x}) = \tfrac{2}{\lambda} \Bigl[ \sqrt{1 + 
\tfrac{\lambda^2 H^2 \ln(a)}{16 \pi^2} } - 1\Bigr] } \nonumber \\
& & \hspace{0.7cm} + \mathcal{A}_0(t,\vec{x}) - \tfrac{\lambda^2 H^3}{32 \pi^2}
\! \int_{0}^{t} \!\!\! dt' \, \mathcal{A}_{0}(t',\vec{x}) + 
\tfrac{\lambda^3 H^3}{64 \pi^2} \! \int_{0}^{t} \!\!\! dt' \, 
[\mathcal{A}_{0}(t',\vec{x})]^2 + O(\lambda^4) \; . \label{Astoch} \qquad
\end{eqnarray}

A number of features of the stochastic solution (\ref{Astoch}) deserve 
comment. First, its expectation value agrees (at leading logarithm order) 
with exact, dimensionally regulated and fully BPHZ renormalized computations 
at 1-loop \cite{Miao:2021gic} and 2-loop \cite{Woodard:2023rqo} orders,
\begin{equation}
\langle \Omega \vert \mathcal{A}(t,\vec{x}) \vert \Omega \rangle = 
\tfrac{\lambda H^2 \ln(a)}{2^4 \pi^2} + \tfrac{\lambda^3 H^5 \ln^2(a)}{
2^{10} \pi^4} + O(\lambda^5) \; . \label{AVEV}
\end{equation}
Second, the stochastic jitter terms on the second line of (\ref{Astoch})
speed up the rate at which the field rolls down its field strength-induced 
effective potential $V_{\rm eff}(A) = -\tfrac{3 H^4}{8\pi^2} \ln(1 + 
\tfrac{\lambda}{2} A)$. This is because it is easier to fluctuate down a
potential than up. Third, significant evolution persists forever on de 
Sitter background. Fourth, a good approximation has been developed for 
generalizing the key relation (\ref{ddprop}) from de Sitter to an arbitrary 
background which has experienced primordial inflation \cite{Kasdagli:2023nzj}. 
Finally, when this generalization is made, one can solve the Langevin 
equation numerically, and the result shows that even late time fields are 
can experience {\it arbitrarily large} effects from what happened during 
primordial inflation \cite{Woodard:2023cqi}.

\subsection{\normalsize Scalar Corrections to Gravity}

Scalar loop corrections to gravity represent another way-station on the 
road to full quantum gravity. Their 1-loop divergences cause 
renormalizaton-induced logarithms of the form (\ref{RGsource}) whose effect 
on the field strength (\ref{MMCWeyl}) and Newtonian potential 
(\ref{MMCNewton}) have already been presented and will be discussed in the
next section. There are also effects from integrating differentiated 
scalars out of the Einstein equation,
\begin{equation}
R_{\mu\nu} - \tfrac12 g_{\mu\nu} R + g_{\mu\nu} \Lambda = 8\pi G \Bigl\{ 
\partial_{\mu} \phi \partial_{\nu} \phi - \tfrac12 g_{\mu\nu} 
g^{\rho\sigma} \partial_{\rho} \phi \partial_{\sigma} \phi \Bigr\} \; . 
\label{STmn}
\end{equation}
Because the propagator of the graviton field $h_{\mu\nu}$ has the same 
logarithmic tail as the nonlinear sigma model scalars of the previous 
subsection, the differentiated scalar $\phi$ in equation (\ref{STmn}) 
needs to be evaluated in the presence of a constant $h_{\mu\nu}$ background.
This means that the conformally rescaled metric $\widetilde{g}_{\mu\nu} =
\eta_{\mu\nu} + \kappa h_{\mu\nu}$ is constant. 

A key insight is that a metric $g_{\mu\nu} = a^2 \widetilde{g}_{\mu\nu}$, 
with constant $\widetilde{g}_{\mu\nu}$, corresponds to de Sitter with a 
different cosmological constant \cite{Miao:2021gic,Miao:2024nsz}. To prove 
this, first compute the affine connection, assuming $\partial_{\rho}
\widetilde{g}_{\mu\nu} = 0$, then show that the Riemann tensor takes the 
locally de Sitter form,
\begin{equation}
\Gamma^{\rho}_{~\mu\nu} = a H \Bigl(\delta^{\rho}_{~\mu} \delta^0_{~\nu} 
\!+\! \delta^{\rho}_{~\nu} \delta^0_{~\mu} \!-\! \widetilde{g}^{\rho 0}
\widetilde{g}_{\mu\nu} \Bigr) \;\; \Longrightarrow \;\;
R^{\rho}_{~\sigma\mu\nu} = -H^2 \widetilde{g}^{00} ( \delta^{\rho}_{~\mu} 
g_{\sigma\nu} - \delta^{\rho}_{~\nu} g_{\sigma\mu} ) \; . \label{dScurv}
\end{equation}
This means that relation (\ref{ddprop}) for the coincident scalar
propagator becomes,
\begin{equation}
\langle \Omega \vert \partial_{\mu} \phi(x) \partial_{\nu} \phi(x) \vert 
\Omega \rangle_{\widetilde{g}} \longrightarrow - \tfrac{3 (H^2 
\widetilde{g}^{00})^2}{32 \pi^2} \, g_{\mu\nu} \; . \label{newddprop}
\end{equation}
It follows that the stress tensor induced by integrating out the scalar in 
a constant $h_{\mu\nu}$ background is \cite{Miao:2024nsz},
\begin{equation}
\partial_{\mu} \phi \partial_{\nu} \phi - \tfrac12 g_{\mu\nu} g^{\rho\sigma} 
\partial_{\rho} \phi \partial_{\sigma} \phi \longrightarrow
+ \tfrac{3 (H^2 \widetilde{g}^{00})^2}{32 \pi^2} \, g_{\mu\nu} \; .
\label{scalarT}
\end{equation}

The scalar-induced stress tensor (\ref{scalarT}) represents a contribution 
to the cosmological constant of $-\tfrac{3}{4 \pi} G (-\widetilde{g}^{00}
H^2)^2$. For de Sitter the result would be just $-\tfrac{3}{4 \pi} G H^4$,
and explicit computations show that making the 1-loop graviton self-energy
conserved requires that precisely this term must be canceled by a combination 
of the Eddington ($R^2$) counterterm (which makes only a finite contribution 
to the 1-point function for de Sitter) plus an extra, finite renormalization 
of the cosmological constant \cite{Tsamis:2023fri}.

The scalar-induced stress tensor (\ref{scalarT}) is not covariantly conserved 
when one relaxes the assumption that $\widetilde{g}_{\mu\nu}$ is constant. 
This is because (\ref{scalarT}) recovers only the leading logarithm result, 
whereas conservation involves sub-leading logarithms. Although correct, this 
does complicate the gravitational field equations, whose classical part is 
conserved. One fix for this problem is to extend (\ref{scalarT}) to a
generally conserved form derived from the variation of,
\begin{equation}
\Delta \mathcal{L} = -\tfrac{R^2 \sqrt{-g}}{768 \pi^2} \times 
\ln(\tfrac{R}{12 H^2}) \; . \label{DeltaL}
\end{equation}
This agrees with (\ref{scalarT}) because the general Ricci scalar,
\begin{equation}
R = -12 \widetilde{g}^{00} H^2 + 6 H a^{-1} \widetilde{g}^{\alpha\beta}
\widetilde{\Gamma}^{0}_{~\alpha\beta} + a^{-2} \widetilde{R} \; , \label{Ricci}
\end{equation}
coincides with $-12 \widetilde{g}^{00} H^2$ for constant 
$\widetilde{g}_{\mu\nu}$.

\subsection{\normalsize Pure Gravity}

It is desirable to express the Lagrangian of gravity (\ref{Einstein}) in
terms of the conformally rescaled fields  \cite{Tsamis:1992xa},
\begin{eqnarray}
\lefteqn{\mathcal{L}_{\rm Einstein} = a^{D-2} \sqrt{-\widetilde{g}} \, 
\widetilde{g}^{\alpha\beta} \widetilde{g}^{\rho\sigma} \widetilde{g}^{\mu\nu} 
\Bigl\{ \tfrac12 h_{\alpha\rho, \mu} h_{\nu\sigma, \beta} - \tfrac12 
h_{\alpha\beta, \rho} h_{\sigma\mu, \nu} } \nonumber \\
& & \hspace{.5cm} + \tfrac14 h_{\alpha\beta, \rho} h_{\mu\nu, \sigma} 
- \tfrac14 h_{\alpha\rho, \mu} h_{\beta\sigma, \nu} \Bigr\}  
+ (\tfrac{D-2}{2}) a^{D-1} H \sqrt{-\widetilde{g}} \, 
\widetilde{g}^{\rho\sigma} \widetilde{g}^{\mu\nu} h_{\rho\sigma, \mu} 
h_{\nu 0} \; . \qquad \label{Einconf}
\end{eqnarray}
To this must be added the gauge fixing function (\ref{LGF}) and the ghost
Lagrangian. The field equation which must be stochastically simplified is
the variation of the total action with respect to $h_{\mu\nu}(x)$.

There are two distinctions from the case of differentiated 
scalars discussed in the previous subsection. First, it is the {\it same} 
field $h_{\mu\nu}$ which provides both the Langevin kinetic term and the 
effective potential from integrating out differentiated gravitons in the
presence of constant $h_{\mu\nu}$. This came up in a nonlinear sigma 
model which has been studied \cite{Miao:2021gic} and the answer is that 
one adds the two contributions from the same term,
\begin{eqnarray}
\lefteqn{(1 + \tfrac{\lambda}{2} \phi) \partial_{\mu} \Bigl[ \sqrt{-g} \, 
g^{\mu\nu} (1 + \tfrac{\lambda}{2} \phi) \partial_{\nu} \phi\Bigr] } 
\nonumber \\
& & \hspace{1cm} \longrightarrow -3 H a^3 (1 + \tfrac{\lambda}{2} \varphi)^2
[\dot{\varphi} - \dot{\varphi}_0] + \tfrac{\lambda}{4} (1 + \tfrac{\lambda}{2}
\varphi) \tfrac{\partial_{\mu} [ \sqrt{-g} \, g^{\mu\nu} \partial_{\nu}
i\Delta(x;x)]}{(1 + \frac{\lambda}{2} \varphi)^2} \; . \qquad
\label{singlefield}
\end{eqnarray}
This procedure reproduces the leading logarithms from explicit
calculations at 1-loop and 2-loop orders, and it should be true generally.

The second issue concerns gauge fixing. The old gauge \cite{Tsamis:1992xa,
Woodard:2004ut},
\begin{equation}
\mathcal{L}_{\rm GF} = -\tfrac12 a^{D-2} \eta^{\mu\nu} F_{\mu} F_{\nu} 
\;\; , \;\; F_{\mu} = \eta^{\rho\sigma} [h_{\mu\rho , \sigma} \!-\!
\tfrac12 h_{\rho\sigma , \mu} \!+\! (D \!-\! 2) a H h_{\mu \rho} 
\delta^{0}_{~\sigma}] \; , \label{oldgauge}
\end{equation}
involves factors of the Minkowski metric $\eta^{\mu\nu}$ and 
$\eta^{\rho\sigma}$ which must be generalized to $\widetilde{g}^{\mu\nu}$
and $\widetilde{g}^{\rho\sigma}$. One must also multiply $\mathcal{L}_{\rm GF}$
by $\sqrt{-\widetilde{g}}$ \cite{Miao:2024shs},
\begin{equation}
\widetilde{\mathcal{L}}_{\rm GF} = -\tfrac12 a^{D-2} \sqrt{-\widetilde{g}} \,
\widetilde{g}^{\mu\nu} \widetilde{F}_{\mu} \widetilde{F}_{\nu} \;\; , \;\; 
\widetilde{F}_{\mu} = \widetilde{g}^{\rho\sigma} [h_{\mu\rho , \sigma} \!-\!
\tfrac12 h_{\rho\sigma , \mu} \!+\! (D \!-\! 2) a H h_{\mu \rho} 
\delta^{0}_{~\sigma}] \; . \label{newgauge}
\end{equation}
The ghost Lagrangian is,
\begin{equation}
\widetilde{\mathcal{L}}_{\rm gh} = -a^{D-2} \sqrt{-\widetilde{g}} \, 
\widetilde{g}^{\mu\nu} \overline{c}_{\mu} \delta \widetilde{F}_{\nu} \; .
\label{ghost}
\end{equation}
Here one must subject all the graviton fields, including those in the factor
of $\widetilde{g}^{\rho\sigma}$ to a linearized general coordinate
transformation, with the infinitesimal parameter replaced by the ghost field
$c^{\alpha}$,
\begin{eqnarray}
\lefteqn{\delta \widetilde{F}_{\nu} = \widetilde{g}^{\rho\sigma} [\delta 
h_{\nu \rho, \sigma} \!-\! \tfrac12 \delta h_{\rho\sigma , \nu} \!-\! 
(D\!-\!2) a \widetilde{H} \delta h_{\nu\rho} u_{\sigma}] } \nonumber \\
& & \hspace{5cm} + \delta \widetilde{g}^{\rho\sigma} [ h_{\nu \rho , \sigma} 
\!-\! \tfrac12 h_{\rho\sigma , \nu} \!-\! (D\!-\!2) a \widetilde{H} 
h_{\nu\rho} u_{\sigma}] \; . \qquad \label{deltaF}
\end{eqnarray}

Expressing the graviton propagator is facilitated by writing the
conformally transformed metric in ADM (Arnowitt, Deser and Misner) form
\cite{Arnowitt:1962hi},
\begin{equation}
\widetilde{g}_{\mu\nu} dx^{\mu} dx^{\nu} \equiv -N^2 d\eta^2 + \gamma_{ij}
(dx^i \!-\! N^i d\eta) (dx^j \!-\! N^j d\eta) \; . \label{ADM}
\end{equation}
The spatial metric can be extended to a full spacetime form,
\begin{eqnarray}
\overline{\gamma}^{\mu\nu} \equiv \left( \begin{matrix} 
0 & 0 \\ 0 & \gamma^{mn} \end{matrix} \right) \;\; , \;\; u^{\mu} \equiv
\left( \begin{matrix} \tfrac1{N} \\ \tfrac{N^m}{N} \end{matrix} \right) 
&\!\!\! \Longrightarrow \!\!\!& \widetilde{g}^{\mu\nu} = 
\overline{\gamma}^{\mu\nu} - u^{\mu} u^{\nu} \; , \qquad 
\label{contravariant} \\
\overline{\gamma}_{\mu\nu} \equiv \left( \begin{matrix} 
\gamma_{k\ell} N^k N^{\ell} & -\gamma_{n\ell} N^{\ell} \\ -\gamma_{mk} N^k 
& \gamma_{mn} \end{matrix} \right) \;\; , \;\; u_{\mu} \equiv
\left( \begin{matrix} -N \\ 0 \end{matrix} \right) 
&\!\!\! \Longrightarrow \!\!\!& \widetilde{g}_{\mu\nu} = 
\overline{\gamma}_{\mu\nu} - u_{\mu} u_{\nu} \; . \qquad \label{covariant}
\end{eqnarray}
Constant $\widetilde{g}_{\mu\nu}$ background corresponds to de Sitter
with Hubble parameter $\widetilde{H} \equiv H/N$. The graviton and ghost
propagators for constant $\widetilde{g}_{\mu\nu}$ are \cite{Miao:2024shs},
\begin{eqnarray}
i[\mbox{}_{\mu\nu} \widetilde{\Delta}_{\rho\sigma}](x;x') &\!\!\! = \!\!\!& 
\sum_{I=A,B,C} [\mbox{}_{\mu\nu} \widetilde{T}^I_{\rho\sigma}] \!\times\! 
i\widetilde{\Delta}_I(x;x') \; , \qquad \label{newgravprop} \\
i[\mbox{}_{\mu} \widetilde{\Delta}_{\rho}](x;x') &\!\!\! = \!\!\!& 
\overline{\gamma}_{\mu\rho} \!\times\! i\widetilde{\Delta}_A(x;x') - 
u_{\mu} u_{\rho} \!\times\! i\widetilde{\Delta}_B(x;x') \; . \qquad 
\label{newghostprop}
\end{eqnarray}
The scalar propagators $i \widetilde{\Delta}_I(x;x')$ are just (\ref{Adef}),
(\ref{Bdef}) and (\ref{Cdef}) with $H$ everywhere replaced by 
$\widetilde{H}$. The various tensor factors are,
\begin{eqnarray}
[\mbox{}_{\mu\nu} \widetilde{T}^A_{\rho\sigma}] & = & 2 
\overline{\gamma}_{\mu (\rho} \overline{\gamma}_{\sigma) \nu} - 
\tfrac{2}{D-3} \overline{\gamma}_{\mu\nu} \overline{\gamma}_{\rho\sigma} 
\qquad , \qquad [\mbox{}_{\mu\nu} \widetilde{T}^B_{\rho\sigma}] = -4 
u_{(\mu} \overline{\gamma}_{\nu) (\rho} u_{\sigma)} \; , \qquad 
\label{newTATB} \\
\, [\mbox{}_{\mu\nu} \widetilde{T}^C_{\rho\sigma}] & = & \tfrac{2}{(D-2) 
(D-3)} [(D\!-\!3) u_{\mu} u_{\nu} \!+\! \overline{\gamma}_{\mu\nu}] 
[(D\!-\!3) u_{\rho} u_{\sigma} \!+\! \overline{\gamma}_{\rho\sigma}]
\; . \qquad \label{newTC}
\end{eqnarray}
The required coincidence limits are,
\begin{eqnarray}
\partial_{\alpha} i [\mbox{}_{\mu\nu} \widetilde{\Delta}_{\rho
\sigma}](x;x')\Bigl\vert_{x'=x} &\!\!\! = \!\!\!& -a \widetilde{H} 
\widetilde{k} u_{\alpha} [\mbox{}_{\mu\nu} \widetilde{T}^A_{\rho\sigma}] 
\; , \qquad \label{1dgravprop} \\
\partial_{\alpha} \partial'_{\beta} i [\mbox{}_{\mu\nu} 
\widetilde{\Delta}_{\rho\sigma}](x;x') \Bigl\vert_{x'=x} &\!\!\! = 
\!\!\!& \nonumber \\
& & \hspace{-1.1cm} \tfrac{\widetilde{k} \widetilde{H}^2}{D} 
g_{\alpha\beta} \Bigl\{ -(D-1) [\mbox{}_{\mu\nu} \widetilde{T}^A_{
\rho\sigma}] + [\mbox{}_{\mu\nu} \widetilde{T}^B_{\rho\sigma}] - 
\tfrac{2}{D-2} [\mbox{}_{\mu\nu} \widetilde{T}^C_{\rho\sigma}] \Bigr\} 
, \qquad \label{2dgravprop} \\
\partial_{\alpha} i [\mbox{}_{\mu} \widetilde{\Delta}_{\rho}](x;x') 
\Bigl\vert_{x'=x} &\!\!\! = \!\!\!& -a \widetilde{H} \widetilde{k} 
u_{\alpha} \overline{\gamma}_{\mu\rho} \; , \qquad \label{1dghostprop} \\
\partial_{\alpha} \partial'_{\beta} i [\mbox{}_{\mu} 
\widetilde{\Delta}_{\rho}](x;x') \Bigl\vert_{x'=x} &\!\!\! = \!\!\!&
-\tfrac{\widetilde{k} \widetilde{H}^2}{D} g_{\alpha\beta} \Bigl\{
(D\!-\!1) \overline{\gamma}_{\mu\rho} + u_{\mu} u_{\rho} \Bigr\} 
\; . \qquad \label{2dghostprop}
\end{eqnarray} 

When the invariant Lagrangian (\ref{Einconf}) and the gauge fixing term
(\ref{newgauge}) are combined there are six terms which can usefully
be combined into three groups \cite{Miao:2024shs},
\begin{equation}
\mathcal{L}_{\rm Einstein} + \widetilde{\mathcal{L}}_{\rm GF} = 
\mathcal{L}_{\scriptscriptstyle 1+2+3} + \mathcal{L}_{\scriptscriptstyle 
4+5} + \mathcal{L}_{\scriptscriptstyle 6} \; . \label{threeterms}
\end{equation}
The three combinations are,
\begin{eqnarray}
\mathcal{L}_{\scriptscriptstyle 1+2+3} &\!\!\! = \!\!\!& a^{D-2} 
\sqrt{-\widetilde{g}} \, \widetilde{g}^{\alpha\beta} [-\tfrac14 
\widetilde{g}^{\gamma\rho} \widetilde{g}^{\delta\sigma} + \tfrac18
\widetilde{g}^{\gamma\delta} \widetilde{g}^{\rho\sigma}] 
h_{\gamma\delta , \alpha} h_{\rho\sigma , \beta} \nonumber \\
& & \hspace{3.5cm} + (\tfrac{D-2}{2}) a^D \sqrt{-\widetilde{g}} \, 
\widetilde{g}^{\gamma\alpha} \widetilde{g}^{\beta\rho} 
\widetilde{g}^{\sigma\delta} h_{\alpha\beta} h_{\rho\sigma} 
\widetilde{H}^2 u_{\gamma} u_{\delta} \; , \qquad \label{L123} \\
\mathcal{L}_{\scriptscriptstyle 4+5} &\!\!\! = \!\!\!& 
\sqrt{-\widetilde{g}} \, \widetilde{g}^{\alpha\beta} 
\widetilde{g}^{\gamma\delta} \widetilde{g}^{\rho\sigma} \partial_{\beta}
\Bigl[ -\tfrac12 \partial_{\sigma} (a^{D-2} h_{\gamma\rho} 
h_{\delta\alpha}) + a^{D-2} h_{\gamma\rho} h_{\delta \alpha , \sigma}
\Bigr] \; , \qquad \label{L45} \\
\mathcal{L}_{\scriptscriptstyle 6} &\!\!\! = \!\!\!& (\tfrac{D-2}2)
\kappa H a^{D-1} \sqrt{-\widetilde{g}} \, \widetilde{g}^{\alpha\beta}
\widetilde{g}^{\gamma\delta} \widetilde{g}^{\rho\sigma} h_{\rho\sigma ,
\gamma} h_{\delta\alpha} h_{\beta 0} \; . \qquad \label{L6}
\end{eqnarray}
The presence of ghost and anti-ghost fields precludes any useful 
simplifications from combining the ghost Lagrangian (\ref{ghost})
with $\mathcal{L}_{\rm Einstein}$ and $\widetilde{\mathcal{L}}_{\rm GF}$.

Three tasks remain:
\begin{enumerate*}
\item{Vary the actions associated with $\widetilde{\mathcal{L}}_{\rm gh}$ 
and (\ref{L123}-\ref{L6}) with respect to $h_{\mu\nu}$;}
\item{Derive an effective stress tensor by integrating out the 
differentiated fields in a constant $\widetilde{g}_{\alpha\beta}$ 
background; and}
\item{Derive the Langevin kinetic operator.}
\end{enumerate*}
The last step involves making a $3+1$ decomposition of the graviton field.
The purely spatial components of $h_{\mu\nu}$ receive a stochastic jitter 
term because their $A$-type propagator possesses a tail (\ref{4Dtail}); 
the mixed space-time and purely temporal components of $h_{\mu\nu}$ have 
no stochastic jitter because their $B$-type and $C$-type propagators
possess no tail. The evolution of these mixed and purely temporal 
components is driven by the evolution of the purely spatial components.
Because this last step has not yet been completed, only the reduction of
$\mathcal{L}_{\rm \scriptscriptstyle 1+2+3}$ will be presented here.

The contribution from (\ref{L123}) to the Euler-Lagrange equation is,
\begin{eqnarray}
\lefteqn{\tfrac{\delta S_{\scriptscriptstyle 1+2+3}}{\delta 
h_{\mu\nu}(x)} } \nonumber \\
& & \hspace{-0.5cm} = \partial_{\alpha} \Bigl\{ a^{D-2} \!\!
\sqrt{-\widetilde{g}} \, \widetilde{g}^{\alpha\beta} [\tfrac12 
\widetilde{g}^{\mu \rho} \widetilde{g}^{\nu \sigma} \!\!\!-\! \tfrac14 
\widetilde{g}^{\mu\nu} \widetilde{g}^{\rho\sigma}] h_{\rho\sigma , \beta}
\Bigr\} \!+\! {\scriptstyle (D - 2)} \widetilde{H}^2 a^D \!\! 
\sqrt{-\widetilde{g}} \, u^{(\mu} \widetilde{g}^{\nu) (\rho} u^{\sigma)} 
h_{\rho\sigma} \nonumber \\
& & \hspace{2cm} - \tfrac{\partial}{\partial h_{\mu\nu}} \Bigl\{a^{D-2} 
\!\! \sqrt{-\widetilde{g}} \, \widetilde{g}^{\gamma\delta} [\tfrac14 
\widetilde{g}^{\alpha \rho} \widetilde{g}^{\beta \sigma} \!\!\!-\! 
\tfrac18 \widetilde{g}^{\alpha\beta} \widetilde{g}^{\rho\sigma}] \Bigr\} 
h_{\alpha\beta , \gamma} h_{\rho\sigma , \delta} \nonumber \\
& & \hspace{3.4cm} + (\tfrac{D-2}{2}) \tfrac{\partial}{\partial h_{\mu\nu}} 
\Bigl\{ a^D \sqrt{-\widetilde{g}} \, \widetilde{g}^{\gamma \alpha} 
\widetilde{g}^{\beta \rho} \widetilde{g}^{\sigma\delta} \Bigr\} 
h_{\alpha\beta} h_{\rho\sigma} \widetilde{H}^2 u_{\gamma} u_{\delta}
. \qquad \label{123EQN}
\end{eqnarray}
The next step is integrating out the differentiated graviton fields
which appear only in the first and third terms of (\ref{123EQN}).   
Because the coincident propagators (\ref{1dgravprop}-\ref{2dgravprop}) 
are finite in dimensional regularization one may as well set $D=4$.
Integrating out the differentiated graviton from the first term gives,
\begin{eqnarray}
\lefteqn{\partial_{\alpha} \Bigl\{ a^2 \!\! \sqrt{-\widetilde{g}} \, 
\widetilde{g}^{\alpha\beta} [\tfrac12 \widetilde{g}^{\mu \rho} 
\widetilde{g}^{\nu \sigma} \!\!\!-\! \tfrac14 \widetilde{g}^{\mu\nu} 
\widetilde{g}^{\rho\sigma}] h_{\rho\sigma , \beta} \Bigr\} }
\nonumber \\
& & \hspace{0.5cm} \longrightarrow -\partial_{\alpha} \Bigl\{ a^3 
\widetilde{H} \widetilde{k} u_{\beta} [\mbox{}_{\gamma\delta} 
\widetilde{T}^{A}_{\rho\sigma}] \times \tfrac{\partial}{\partial 
h_{\gamma\delta}} \Bigl[ \sqrt{-\widetilde{g}} \, 
\widetilde{g}^{\alpha\beta} [\tfrac12 \widetilde{g}^{\mu\rho}
\widetilde{g}^{\nu\sigma} \!-\! \tfrac14 \widetilde{g}^{\mu\nu}
\widetilde{g}^{\rho\sigma}] \Bigr] \Bigr\} \; , \qquad \\
& & \hspace{0.5cm} \longrightarrow a^4 \kappa \widetilde{H}^2
\widetilde{k} \sqrt{-\widetilde{g}} \, [\tfrac{33}2 
\overline{\gamma}^{\mu\nu} - \tfrac92 \widetilde{g}^{\mu\nu}] \; . 
\qquad \label{term1} 
\end{eqnarray}
The result for the third term of (\ref{123EQN}) is,
\begin{eqnarray}
\lefteqn{ - \tfrac{\partial}{\partial h_{\mu\nu}} \Bigl\{a^{2} 
\!\! \sqrt{-\widetilde{g}} \, \widetilde{g}^{\gamma\delta} [\tfrac14 
\widetilde{g}^{\alpha \rho} \widetilde{g}^{\beta \sigma} \!\!\!-\! 
\tfrac18 \widetilde{g}^{\alpha\beta} \widetilde{g}^{\rho\sigma}] \Bigr\} 
h_{\alpha\beta , \gamma} h_{\rho\sigma , \delta} } \nonumber \\
& & \hspace{0cm} \longrightarrow \tfrac{\partial}{\partial 
h_{\mu\nu}} \Bigl\{a^{2} \!\! \sqrt{-\widetilde{g}} \, 
\widetilde{g}^{\gamma\delta} [\tfrac14 \widetilde{g}^{\alpha \rho} 
\widetilde{g}^{\beta \sigma} \!\!\!-\! \tfrac18 
\widetilde{g}^{\alpha\beta} \widetilde{g}^{\rho\sigma}] \Bigr\} 
\nonumber \\
& & \hspace{5cm} \times \tfrac{\widetilde{k} \widetilde{H}^2}{4} 
g_{\gamma\delta} \Bigl\{ 3 [\mbox{}_{\alpha\beta} 
\widetilde{T}^A_{\rho\sigma}] - [\mbox{}_{\alpha\beta} 
\widetilde{T}^B_{\rho\sigma}] + [\mbox{}_{\alpha\beta}
\widetilde{T}^C_{\rho\sigma}] \Bigr\} , \qquad \\
& & \hspace{0cm} \longrightarrow a^4 \kappa \widetilde{H}^2 
\widetilde{k} \sqrt{-\widetilde{g}} \, [-\tfrac72 \overline{\gamma}^{\mu\nu} 
- \tfrac52 u^{\mu} u^{\nu}] \; . \qquad \label{term3}
\end{eqnarray}

The last step is to decompose the graviton field into components which
possess a tail, and hence a stochastic jitter term, and those which do not.
The appropriate representation seems to be,
\begin{equation}
\kappa h_{\mu\nu} \equiv A_{\mu\nu} + 2 u_{(\mu} B_{\nu)} + [u_{\mu} u_{\nu} +
\overline{\gamma}_{\mu\nu}] C \qquad , \qquad u^{\alpha} A_{\alpha\beta} = 0
= u^{\alpha} B_{\alpha} \; . \label{components}
\end{equation}
The first two terms of (\ref{123EQN}) represent the action of the 
gauge-fixed kinetic operator on the graviton field,
\begin{eqnarray}
\lefteqn{\partial_{\alpha} \Bigl\{ a^{2} \! \sqrt{-\widetilde{g}} \, 
\widetilde{g}^{\alpha\beta} [\tfrac12 \widetilde{g}^{\mu \rho} 
\widetilde{g}^{\nu \sigma} \!\!-\! \tfrac14 \widetilde{g}^{\mu\nu} 
\widetilde{g}^{\rho\sigma}] \kappa h_{\rho\sigma , \beta} \Bigr\} \!+\! 
2 \widetilde{H}^2 a^4 \! \sqrt{-\widetilde{g}} \, u^{(\mu} 
\widetilde{g}^{\nu) (\rho} u^{\sigma)} \kappa h_{\rho\sigma} } 
\nonumber \\
& & \hspace{1.3cm} \longrightarrow [\tfrac12 \widetilde{g}^{\mu\rho}
\widetilde{g}^{\nu\sigma} - \tfrac14 \widetilde{g}^{\mu\nu} 
\widetilde{g}^{\rho\sigma} ] \widetilde{D}_A A_{\rho\sigma} +
u^{(\mu} \widetilde{g}^{\nu) \rho} \widetilde{D}_B B_{\rho} +
u^{\mu} u^{\nu} \widetilde{D}_C C . \qquad \label{Lang12}
\end{eqnarray}
It is useful to restore $D$ dimensions to define the operators
$\widetilde{D}_{I}$ and their stochastic reductions,
\begin{eqnarray}
\widetilde{D}_A A_{\rho\sigma} &\!\!\! \equiv \!\!\!& \partial_{\alpha}
[a^{D-2} \!\! \sqrt{-\widetilde{g}} \, \widetilde{g}^{\alpha\beta} 
\! A_{\rho\sigma , \beta}] \longrightarrow -{\scriptstyle (D-2)} 
\widetilde{H} a^{D-1} \!\! \sqrt{-\widetilde{g}} \,  u^{\alpha} 
\partial_{\alpha} (A_{\rho\sigma} \!\!-\! a_{\rho\sigma}) , \qquad 
\label{DAdef} \\
\widetilde{D}_B B_{\rho} &\!\!\! \equiv \!\!\!& [\widetilde{D}_A -
{\scriptstyle (D-2)} \widetilde{H}^2 a^D \!\! \sqrt{-\widetilde{g}} \, ]
B_{\rho} \longrightarrow -{\scriptstyle (D-2)} \widetilde{H}^2 a^D
\!\! \sqrt{-\widetilde{g}} \, B_{\rho} \; , \label{DBdef} \\
\widetilde{D}_C C &\!\!\! \equiv \!\!\!& [\widetilde{D}_A -
{\scriptstyle 2 (D-3)} \widetilde{H}^2 a^D \!\! \sqrt{-\widetilde{g}} 
\, ] C \longrightarrow -{\scriptstyle 2 (D-3)} \widetilde{H}^2 a^D
\!\! \sqrt{-\widetilde{g}} \, C \; , \label{DCdef}
\end{eqnarray}
where $a_{\rho\sigma}$ is the stochastic jitter for $A_{\rho\sigma}$.
Putting everything together in $D=4$ gives the final form,
\begin{eqnarray}
\lefteqn{\partial_{\alpha} \Bigl\{ a^{2} \! \sqrt{-\widetilde{g}} \, 
\widetilde{g}^{\alpha\beta} [\tfrac12 \widetilde{g}^{\mu \rho} 
\widetilde{g}^{\nu \sigma} \!\!-\! \tfrac14 \widetilde{g}^{\mu\nu} 
\widetilde{g}^{\rho\sigma}] \kappa h_{\rho\sigma , \beta} \Bigr\} \!+\! 
2 \widetilde{H}^2 a^4 \! \sqrt{-\widetilde{g}} \, u^{(\mu} 
\widetilde{g}^{\nu) (\rho} u^{\sigma)} \kappa h_{\rho\sigma} } 
\nonumber \\
& & \hspace{2cm} \longrightarrow -[\widetilde{g}^{\mu \rho} 
\widetilde{g}^{\nu \sigma} \!-\! \tfrac12 \widetilde{g}^{\mu\nu} 
\widetilde{g}^{\rho\sigma}] \widetilde{H} a^3 \sqrt{-\widetilde{g}}
\, u^{\alpha} \partial_{\alpha} (A_{\rho\sigma} \!-\! a_{\rho\sigma})
\nonumber \\
& & \hspace{6.4cm} - 2 \widetilde{H}^2 a^4 \sqrt{-\widetilde{g}} \, 
[ u^{(\mu} B^{\nu)} \!+\! u^{\mu} u^{\nu} C] . \qquad 
\label{Lang12final}
\end{eqnarray}
The third term of equation (\ref{123EQN}) makes no contribution to the
Langevin kinetic operator because it is the product of two first 
derivatives. In contrast, the final term of (\ref{123EQN}) contributes
without any change although one should take the derivative with respect
to $h_{\mu\nu}$,
\begin{eqnarray} 
\lefteqn{ \tfrac{\partial}{\partial h_{\mu\nu}} 
\Bigl\{ a^D \sqrt{-\widetilde{g}} \, \widetilde{g}^{\gamma \alpha} 
\widetilde{g}^{\beta \rho} \widetilde{g}^{\sigma\delta} \Bigr\} 
h_{\alpha\beta} h_{\rho\sigma} \widetilde{H}^2 u_{\gamma} u_{\delta} }
\nonumber \\
& & \hspace{-0.1cm} = \kappa \widetilde{H}^2 a^4 \!\! \sqrt{-\widetilde{g}} 
\, \Bigl\{ \tfrac12 \widetilde{g}^{\mu\nu} u^{\alpha} \widetilde{g}^{\beta
\rho} u^{\sigma} \!-\! 2 u^{(\mu} \widetilde{g}^{\nu) \alpha}
\widetilde{g}^{\beta \rho} u^{\sigma} \!-\! u^{\alpha} \widetilde{g}^{\beta
(\mu} \widetilde{g}^{\nu) \rho} u^{\sigma} \Bigr\} h_{\alpha\beta}
h_{\rho\sigma} . \qquad \label{term4}
\end{eqnarray}

The full reduction of (\ref{123EQN}) is the combination of expressions
(\ref{term1}), (\ref{term3}), (\ref{Lang12final}) and (\ref{term4}). 
It will simplify the result if we multiply by a factor of $\kappa a^{-4}/
\sqrt{-\widetilde{g}}$,
\begin{eqnarray}
\lefteqn{ \tfrac{\kappa a^{-4}}{\sqrt{-\widetilde{g}}} \times 
\tfrac{\delta S_{1+2+3}}{\delta h_{\mu\nu}(x)} \longrightarrow
\tfrac{\kappa^2 \widetilde{H}^4}{8 \pi^2} \, [13 \widetilde{g}^{\mu\nu}
+ 6 u^{\mu} u^{\nu}] } \nonumber \\
& & \hspace{0cm} - [\widetilde{g}^{\mu \rho} \widetilde{g}^{\nu \sigma} 
\!-\! \tfrac12 \widetilde{g}^{\mu\nu} \widetilde{g}^{\rho\sigma}] a^{-1}
\widetilde{H} u^{\alpha} \partial_{\alpha} (A_{\rho\sigma} \!-\! 
a_{\rho\sigma}) - 2 \widetilde{H}^2 [ u^{(\mu} B^{\nu)} \!+\! u^{\mu}
u^{\nu} C] \nonumber \\
& & \hspace{1cm} + \kappa^2 \widetilde{H}^2 \Bigl\{ \tfrac12 
\widetilde{g}^{\mu\nu} u^{\alpha} \widetilde{g}^{\beta\rho} u^{\sigma} 
\!-\! 2 u^{(\mu} \widetilde{g}^{\nu) \alpha} \widetilde{g}^{\beta \rho} 
u^{\sigma} \!-\! u^{\alpha} \widetilde{g}^{\beta(\mu} 
\widetilde{g}^{\nu) \rho} u^{\sigma} \Bigr\} h_{\alpha\beta} 
h_{\rho\sigma} . \qquad \label{123REDUCED}
\end{eqnarray}
Steps 1 and 2 have been implemented on the other three Lagrangians
(\ref{ghost}), (\ref{L45}) and (\ref{L6}) \cite{Miao:2024shs}, but the 
final step has not yet been completed. The graviton-induced stress 
tensor which results from step 2 must also be checked against a
dimensionally regulated and renormalized computation of the graviton
1-point function, the same as was done for scalar loop contributions
to gravity \cite{Tsamis:2023fri,Miao:2024nsz}. Note that it is not 
possible to use an existing result in the old gauge \cite{Tsamis:2005je};
because $N$-point functions are gauge dependent one must re-do the 
computation in the new gauge (\ref{newgauge}).

The fully reduced equation would suffice to compute stochastic 
corrections to the cosmological background. The graviton mode 
function seems to reside in the field $A_{ij}$ so one might also
check for stochastic enhancements of it. The double logarithmic 
corrections (\ref{GRmode}) that were found in the old gauge 
\cite{Tan:2021lza} could derive from a contribution to the 
stochastic Langevin equation which does seem to be present in
(\ref{123REDUCED}),
\begin{equation}
\widetilde{H} a^{-1} u^{\alpha} \partial_{\alpha} (A - a) \sim 
\kappa^2 \widetilde{H}^4 A^3 + \dots
\end{equation}
However, checking the numerical coefficient would require a 
dimensionally regulated and fully renormalized computation of the 
graviton loop contribution to $-i[\mbox{}^{\mu\nu} 
\Sigma^{\rho\sigma}](x;x')$ in the new gauge (\ref{newgauge}). Note 
also that the fully reduced stochastic equation will not allow one 
to derive corrections to the gravitational response to a point mass 
such as (\ref{GRNewton}-\ref{GRSlip}) because spatial derivatives 
were dropped in the third step of finding the Langevin kinetic 
operator. The correct procedure might be a straightforward as not 
making the final simplification of dropping the derivatives in 
relations (\ref{DBdef}-\ref{DCdef}).

\section{Modified Renormalization Group}

The point of this section is to explain how to resum large loop
corrections from the second source (\ref{RGsource}) using a variant 
of the renormalization group in which certain combinations of BPHZ 
counterterms are regarded as curvature-dependent renormalizations of 
parameters in the bare Lagrangian. The section begins with a 
discussion of how this works for a simple nonlinear sigma model. The
technique is then generalized to the case of matter loop corrections 
to gravity, which all require the same two 1-loop counterterms 
(\ref{DeltaEinstein}). The section closes with a discussion of graviton 
loop corrections to matter, which each possesses its own set of 
counterterms.

\subsection{\normalsize Nonlinear Sigma Models}

In addition to the stochastic effects discussed in section 3.2, the
2-field nonlinear sigma model (\ref{LAB}) experiences large logarithms 
induced by the incomplete cancellation between primitive divergences 
and counterterms as in expression (\ref{RGsource}). These manifest in 
a number of ways, including the $B$ exchange potential. Adding a point
source $J(t,\vec{x}) = K a(t) \delta^3(\vec{x})$ to the linearized 
effective field equation for the field $B$ and taking the late time 
limit results in a solution which involves powers of $\ln(H r)$
\cite{Miao:2021gic},
\begin{equation}
B(t,r) \longrightarrow \tfrac{K H \ln(Hr)}{4\pi} \Bigl\{1 - 
\tfrac{\lambda^2 H^2 \ln(H r)}{32 \pi^2} + O(\lambda^4) \Bigr\} \; . 
\label{Bpotential}
\end{equation} 
These can be explained using a variant of the renormalization group. 
The 1-loop self-mass for $B$ requires two higher derivative counterterms,
\begin{equation}
\Delta \mathcal{L} = -\tfrac12 C_{B1} \square B \square B \sqrt{-g} -
\tfrac12 C_{B2} R \partial_{\mu} B \partial_{\nu} B g^{\mu\nu} \sqrt{-g}
\; . \label{cterms}
\end{equation}
The one proportional to $C_{B2}$ can be viewed as a curvature-dependent
field strength renormalization whose associated $\gamma$ function is
\cite{Miao:2021gic},
\begin{equation}
Z_B = 1 + C_{B2} \!\times\! R + O(\lambda^4) \quad \Longrightarrow \quad 
\gamma_B \equiv \tfrac{\partial \ln(Z_B)}{\partial \ln(\mu^2)} = -
\tfrac{\lambda^2 H^2}{32 \pi^2} + O(\lambda^4) \; . \label{gammaB}
\end{equation}
The $B$ exchange potential $\Phi_B(t,r)$ is a 2-point Green's function 
so the Callen-Symanzik equation for it reads,
\begin{equation}
\Bigl[ \tfrac{\partial}{\partial \ln(\mu)} + \beta 
\tfrac{\partial}{\partial \lambda} + 2 \gamma_{B} \Bigr] B(t,r) = 0 \; .
\label{CSeqn}
\end{equation}
Recognizing that the $\beta$-function vanishes \cite{Woodard:2023rqo}, 
and making the replacement $\ln(\mu) \rightarrow -\ln(Hr)$, not only
explains (\ref{Bpotential}) but also permits a resummation,
\begin{equation}
B(t,r) \longrightarrow \tfrac{K H \ln(H r)}{4\pi} \times [H r
]^{-\frac{\lambda^2 H^2}{32 \pi^2}} \; . \label{fullBpot}
\end{equation}

\subsection{\normalsize Matter Corrections to Gravity}

All 1-loop matter corrections to gravity require the same two counterterms
(\ref{DeltaEinstein}), which is repeated here,
\begin{equation}
\Delta \mathcal{L}_{\rm Einstein} = c_1 R^2 \sqrt{-g} + c_2 
C^{\alpha\beta\gamma\delta} C_{\alpha\beta\gamma\delta} \sqrt{-g} \; .
\label{newDEinstein}
\end{equation}
Classical $D$-dimensional gravity with a cosmological constant $\Lambda$ 
obeys $R_{\mu\nu} = \Lambda g_{\mu\nu}$ and $R = D \Lambda$. It therefore
makes sense to expand the Eddington ($R^2$) counterterm,
\begin{equation}
R^2 = (R - D \Lambda)^2 + 2 D \Lambda [R - (D\!-\!2) \Lambda] + D
(D\!-\!4) \Lambda^2 \; . \label{Rsqexpand}
\end{equation}
The combination $(R - D \Lambda)^2$ is a genuine higher derivative
counterterm, analogous to $\square B \square B$ in expression 
(\ref{cterms}). It plays no role in explaining the leading logarithm
corrections from matter loops. However, $R - (D-2) \Lambda$ is $\kappa^2$
times the Einstein-Hilbert Lagrangian (\ref{Einstein}), and it could be
regarded as a $\Lambda$-dependent field strength renormalization of the
graviton field \cite{Miao:2024nsz}. The final contribution ($D (D-4)
\Lambda^2$) is not divergent owing to the factor of $(D-4)$.

A similar expansion of the Weyl ($C^{\alpha\beta\gamma\delta}
C_{\alpha\beta\gamma\delta}$) counterterm can be made by using the fact
that the Gauss-Bonnet scalar (times $\sqrt{-g}$) is a total derivative in 
$D=4$ dimensions,
\begin{equation}
G \equiv R^{\mu\nu\rho\sigma} R_{\mu\nu\rho\sigma} - 4 R^{\mu\nu} 
R_{\mu\nu} + R^2 \; . \label{GaussB}
\end{equation}
The $D$-dimensional Weyl tensor obeys,
\begin{equation}
C^{\mu\nu\rho\sigma} C_{\mu\nu\rho\sigma} = R^{\mu\nu\rho\sigma}
R_{\mu\nu\rho\sigma} - \tfrac{4}{D-2} R^{\mu\nu} R_{\mu\nu} 
+ \tfrac{2}{(D-1) (D-2)} R^2 \; . \label{Csquared}
\end{equation}
Hence the Weyl counterterm can be written as,
\begin{equation}
C^{\mu\nu\rho\sigma} C_{\mu\nu\rho\sigma} \longrightarrow 2 R^{\mu\nu} 
R_{\mu\nu} - \tfrac23 R^2 \; . \label{Weylcterm}
\end{equation}
The square of the Ricci tensor can be expanded analogous to 
(\ref{Rsqexpand}),
\begin{equation}
R^{\mu\nu} R_{\mu\nu} = (R^{\mu\nu} - \Lambda g^{\mu\nu}) (R_{\mu\nu}
- g_{\mu\nu} \Lambda) + 2 D \Lambda [R - (D\!-\!2) \Lambda] + (D\!-\!4)
\Lambda^2 \; . \label{Riccitensor}
\end{equation}
Combining this with (\ref{Rsqexpand}) and (\ref{Csquared}) implies,
\begin{equation}
C^{\mu\nu\rho\sigma} C_{\mu\nu\rho\sigma} \longrightarrow 2 (R^{\mu\nu}
\!-\! g^{\mu\nu} \Lambda) (R_{\mu\nu} \!-\! g_{\mu\nu} \Lambda) -
\tfrac23 (R \!-\! 4 \Lambda)^2 - \tfrac43 \Lambda [R \!-\! 2 \Lambda] \; .
\label{Csqexpand}
\end{equation}

Expressions (\ref{Rsqexpand}) and (\ref{Csqexpand}) imply that the
counterterm (\ref{DeltaEinstein}) can be regarded as some irrelevant
higher derivative counterterms plus a graviton field strength 
renormalization. The associated gamma function is,
\begin{equation}
\delta Z = (D\!-\!1) \Bigl[ 2 D c_1 \!-\! \tfrac43 c_2\Bigr] \kappa^2
H^2 \qquad \Longrightarrow \qquad \gamma = 
\tfrac{\partial \ln(1 + \delta Z)}{\partial \ln(\mu^2)} \; .
\end{equation}
The Weyl tensor for gravitational radiation and the potentials can be
viewed as 2-point Green's functions for which the Callan-Symanzik equation
implies,
\begin{equation}
\Bigl[ \tfrac{\partial}{\partial \ln(\mu)} + 2 \gamma \Bigr] G^{(2)} = 0 
\; . \label{CallanSEQN}
\end{equation}
The renormalization scale $\ln(\mu)$ only enters in the form $\ln(\mu a)$ 
through relation (\ref{RGsource}) so one can replace derivatives with 
respect to $\ln(\mu)$ by derivatives with respect to $\ln(a)$. The only 
thing that changes from one matter theory to another is what the 
coefficients $c_1$ and $c_2$ are.

\subsubsection{Einstein-MMC Scalar}

The coefficients $c_1$ and $c_2$ for a massless, minimally coupled scalar
are \cite{Park:2011ww,Miao:2024atw},
\begin{equation}
c_1 = \tfrac1{2^7 \cdot 3 \cdot \pi^2} \times \tfrac{\mu^{D-4}}{D-4} 
\;\; , \;\; c_2 = \tfrac1{2^7 \cdot 3 \cdot 5 \cdot \pi^2} \times 
\tfrac{\mu^{D-4}}{D-4} \qquad \Longrightarrow \qquad \gamma = 
\tfrac{3 \kappa^2 H^2}{320 \pi^2} \; . \label{MMCgamma}
\end{equation}
These results explain the large logarithms in the 1-loop corrections 
(\ref{MMCWeyl}) to the Weyl tensor and (\ref{MMCNewton}) to the Newtonian 
potential. These results can even be resumed as long as the de Sitter 
phase persists,
\begin{eqnarray}
C_{0i0j} &\!\!\! \longrightarrow \!\!\!& C_{0i0j}^{\rm tree} \times
[a(t)]^{-\frac{3 \kappa^2 H^2}{160 \pi^2}} \; , \qquad 
\label{MMCWeylsum} \\
\Psi &\!\!\! \longrightarrow \!\!\!& \tfrac{G M}{a r} \times [a(t) H r
]^{-\frac{3 \kappa^2 H^2}{160 \pi^2}} \; . \qquad \label{MMCNewtonsum}
\end{eqnarray}
The renormalization group does not explain the 1-loop results 
(\ref{MMCSlip}) for the gravitational slip because they are not leading 
logarithm.

\subsubsection{Einstein-MCC Scalar}

The coefficients $c_1$ and $c_2$ for a massless, minimally coupled scalar
are \cite{Capper:1973bk},
\begin{equation}
c_1 = 0 \;\; , \;\; c_2 = \tfrac1{2^7 \cdot 3 \cdot 5 \cdot \pi^2} \times 
\tfrac{\mu^{D-4}}{D-4} \qquad \Longrightarrow \qquad \gamma = 
-\tfrac{\kappa^2 H^2}{960 \pi^2} \; . \label{MCCgamma}
\end{equation}
The renormalization group explains (\ref{MCCWeyl}-\ref{MCCNewton}). It also
implies fully resummed results \cite{Foraci:2024cwi},
\begin{eqnarray}
C_{0i0j} &\!\!\! \longrightarrow \!\!\!& C_{0i0j}^{\rm tree} \times
[a(t)]^{\frac{\kappa^2 H^2}{480 \pi^2}} \; , \qquad 
\label{MCCWeylsum} \\
\Psi &\!\!\! \longrightarrow \!\!\!& \tfrac{G M}{a r} \times [a(t) H r
]^{\frac{\kappa^2 H^2}{480 \pi^2}} \; . \qquad \label{MCCNewtonsum}
\end{eqnarray}

\subsubsection{Einstein-Dirac}

It turns out that a loop of any conformally invariant field is proportional 
to that of the massless, conformally coupled scalar \cite{Duff:2000mt}. The
factor of a loop of massless, Dirac fermions is $6$ \cite{Capper:1973mv},
\begin{equation}
c_1 = 0 \;\; , \;\; c_2 = \tfrac1{2^6 \cdot 5 \cdot \pi^2} \times 
\tfrac{\mu^{D-4}}{D-4} \qquad \Longrightarrow \qquad \gamma = 
-\tfrac{\kappa^2 H^2}{160 \pi^2} \; . \label{Diracgamma}
\end{equation}
It follows that the renormalization group explains both the Weyl tensor
(\ref{DiracWeyl}) and the Newtonian potential (\ref{DiracNewton}), and
these results can be resummed to give \cite{Foraci:2024cwi},
\begin{eqnarray}
C_{0i0j} &\!\!\! \longrightarrow \!\!\!& C_{0i0j}^{\rm tree} \times
[a(t)]^{\frac{\kappa^2 H^2}{80 \pi^2}} \; , \qquad 
\label{DiracWeylsum} \\
\Psi &\!\!\! \longrightarrow \!\!\!& \tfrac{G M}{a r} \times [a(t) H r
]^{\frac{\kappa^2 H^2}{80 \pi^2}} \; . \qquad \label{DiracNewtonsum}
\end{eqnarray}

\subsubsection{Einstein-Maxwell}

Electromagnetism is also conformally invariant in $D=4$ dimensions.
A loop of photons contributes a factor of 12 times that of a loop of
massless, conformally coupled scalars \cite{Duff:2000mt,Capper:1974ed},
\begin{equation}
c_1 = 0 \;\; , \;\; c_2 = \tfrac1{2^2 \cdot 5 \cdot \pi^2} \times 
\tfrac{\mu^{D-4}}{D-4} \qquad \Longrightarrow \qquad \gamma = 
-\tfrac{\kappa^2 H^2}{80 \pi^2} \; . \label{EMgamma}
\end{equation}
The corresponding renormalization group analysis not only explains
the 1-loop corrections (\ref{EMWeyl}) and (\ref{EMNewton}) but also 
permits the derivation of resummed results \cite{Foraci:2024vng},
\begin{eqnarray}
C_{0i0j} &\!\!\! \longrightarrow \!\!\!& C_{0i0j}^{\rm tree} \times
[a(t)]^{\frac{\kappa^2 H^2}{40 \pi^2}} \; , \qquad 
\label{EMWeylsum} \\
\Psi &\!\!\! \longrightarrow \!\!\!& \tfrac{G M}{a r} \times [a(t) H r
]^{\frac{\kappa^2 H^2}{40 \pi^2}} \; . \qquad \label{EMNewtonsum}
\end{eqnarray}

\subsection{\normalsize Graviton Corrections to Matter}

Graviton corrections to matter are not as generic as matter loop corrections
to gravity. One reason is that each matter 1PI 2-point function requires its
own set of counterterms. Another reason is that some of these counterterms 
are not general coordinate invariant, owing to the breaking of de Sitter
invariance in the graviton gauge \cite{Tsamis:1992xa,Woodard:2004ut} discussed
in section 2.2. Finally, the counterterms of each matter theory require a
distinct decomposition into those which can be regarded as renormalizations of
bare parameters and those which cannot.

\subsubsection{Einstein-MMC Scalar}

Recall from (\ref{DMMCS}) that graviton corrections to the massless, minimally
coupled scalar self-mass require three counterterms,
\begin{equation}
\Delta \mathcal{L}_{\rm MMC} = -\tfrac{\alpha_1}{2} \square \phi \square 
\phi \sqrt{-g} - \tfrac{\alpha_2}{2} R \partial_{\mu} \phi \partial_{\nu} \phi 
g^{\mu\nu} \sqrt{-g} - \tfrac{\alpha_3}{2} R \partial_{0} \phi \partial_{0}
\phi g^{00} \sqrt{-g} \; . \label{newDeltaMMC}
\end{equation}
The coefficients were found to be \cite{Glavan:2021adm},
\begin{equation}
\alpha_1 = 0 \quad , \quad \alpha_2 R = -\tfrac{\kappa^2 H^2}{4 \pi^2} \times
\tfrac{\mu^{D-4}}{D - 4} \quad , \quad \alpha_3 R = \tfrac{\kappa^2 H^2}{2 \pi^2}
\times \tfrac{\mu^{D-4}}{D-4} \; . \label{MMCcoefs}
\end{equation}
The $\alpha_1$ counterterm is a higher derivative like that of the $C_{B1}$ term 
in (\ref{cterms}) and anyway vanishes. The $\alpha_3$ counterterm arises from the 
de Sitter breaking gauge \cite{Tsamis:1992xa,Woodard:2004ut}. In contrast, one 
can think of $\alpha_2 R$ as a field strength renormalization which implies
\cite{Glavan:2021adm},
\begin{equation}
\delta Z = \alpha_2 R \qquad \Longrightarrow \qquad \gamma \equiv
\tfrac{\partial \ln(1 + \delta Z)}{\partial \ln(\mu^2)} = 
-\tfrac{\kappa^2 H^2}{8 \pi^2} \; . \label{gammaMMC}
\end{equation}
There does not seem to be a renormalization group explanation for the slower 
fall-off of the scalar mode function (\ref{scalarmode}), but the exchange 
potential (\ref{latepot}) follows from the Callan-Symanzik equation if one 
replaces $\ln(\mu)$ with $-\ln(H r)$,
\begin{equation}
\Bigl[ \tfrac{\partial}{\partial \ln(\mu)} + 2 \gamma \Bigr] G^{(2)} = 0 \; .
\label{CSEQN}
\end{equation}
The full resummation is,
\begin{equation}
\phi(t,r) \longrightarrow \tfrac{K H}{4\pi} \ln(H r) \times 
[ H r]^{-\frac{\kappa^2 H^2}{8 \pi^2}} \; . \label{fullMMCpot}
\end{equation}
Note the close analogy between the nonlinear sigma model relations 
(\ref{Bpotential}), (\ref{gammaB}) and (\ref{fullBpot}) and the MMC scalar
relations (\ref{latepot}), (\ref{gammaMMC}) and (\ref{fullMMCpot}).

\subsubsection{Einstein-MCC Scalar}

Graviton corrections to the self-mass of a massless, conformally coupled
scalar require three counterterms (\ref{DeltaMCCconf}) \cite{Glavan:2020gal}, 
\begin{equation}
\Delta \mathcal{L}_{\rm MMC} = -\tfrac{\alpha}{2 a^2} (\partial^2 
\widetilde{\phi})^2 + \tfrac{\beta H^2}{2} \partial_{\mu} \widetilde{\phi} \,
\partial^{\mu} \widetilde{\phi} - \tfrac{\gamma H^2}{2} \partial_i
\widetilde{\phi} \, \partial_i \widetilde{\phi} \; . \label{newDeltaMCCconf}  
\end{equation}
Note that these are expressed on de Sitter background in terms of the 
conformally rescaled field $\widetilde{\phi} \equiv \phi/a^{\frac{D}2 - 1}$.
The coefficients are \cite{Glavan:2020gal},
\begin{equation}
\alpha = -\tfrac{\kappa^2}{48 \pi^2} \times \tfrac{\mu^{D-4}}{D-4} \quad , 
\quad \beta H^2 = \tfrac{19 \kappa^2 H^2}{48 \pi^2} \times \tfrac{\mu^{D-4}}{D-4}
\quad , \quad \gamma H^2 = \tfrac{5 \kappa^2 H^2}{8 \pi^2} \times 
\tfrac{\mu^{D-4}}{D-4} \; . \label{MCCcoefs}
\end{equation}
The gamma function needed for the Callan-Symanzik equation (\ref{CSEQN}) ---
with $\ln(\mu)$ replaced by $\ln(a)$ --- to explain the scalar mode function 
(\ref{MCCmode}) and exchange potential (\ref{MCCphi}) is,
\begin{equation}
{\rm Needed} \qquad \gamma = -\tfrac{\kappa^2 H^2}{48 \pi^2} \qquad 
\Longrightarrow \qquad \delta Z = -\tfrac{\kappa^2 H^2}{24 \pi^2} \times 
\tfrac{\mu^{D-4}}{D-4} \; . \label{gammaMCC}
\end{equation}
There are ways to combine the counterterms to produce this, for example,
\begin{equation}
-2 \alpha H^2 - \beta H^2 + \tfrac12 \gamma H^2 = -\tfrac{\kappa^2 H^2}{24 \pi^2}
\times \tfrac{\mu^{D-4}}{D-4} \; . \label{forexample}
\end{equation}
However, these all seem tendentious and are not supported by the way factors
of $\ln(a)$ appear in the self-mass \cite{Glavan:2020ccz}.

Part of the problem may be that the MCC scalar field equation contains 
differentiated graviton fields through the conformal coupling,
\begin{equation}
\widetilde{R} \sqrt{-\widetilde{g}} = \partial_{\mu} \Bigl[ \sqrt{-\widetilde{g}}
\, ( \widetilde{g}^{\rho\sigma} \widetilde{\Gamma}^{\mu}_{~\rho\sigma} -
\widetilde{g}^{\mu\nu} \widetilde{\Gamma}^{\rho}_{~\rho\nu})\Bigr] +
\sqrt{-\widetilde{g}} \, \widetilde{g}^{\mu\nu} (\widetilde{\Gamma}^{\rho}_{~\mu
\sigma} \widetilde{\Gamma}^{\sigma}_{~\nu\rho} - \widetilde{\Gamma}^{\rho}_{~
\rho\sigma} \widetilde{\Gamma}^{\sigma}_{~\mu\nu}) \; . \label{Riccitilde}
\end{equation}
These should probably be integrated out the same way that differentiated 
gravitons were integrated out of the pure gravitational field equations in
section 3.4. When this is done, and the undifferentiated gravitons are set to 
zero, the result is,  
\begin{equation}
\tfrac{\delta S}{\delta \widetilde{\phi}} = \partial_{\mu} [\sqrt{-\widetilde{g}}
\, \widetilde{g}^{\mu\nu} \partial_{\nu} \widetilde{\phi}] - \tfrac14 
(\tfrac{D-2}{D-1}) \widetilde{\phi} \widetilde{R} \sqrt{-\widetilde{g}} 
\longrightarrow \partial^2 \widetilde{\phi} - \tfrac{\kappa^2 H^4 a^2}{8 \pi^2}
\widetilde{\phi} + \dots \label{MCCEQN}
\end{equation}
However, one would need an extra factor of $+\frac{\kappa^2 H^4 a^2}{12 \pi^2}
\widetilde{\phi}$ in order to explain the graviton loop corrections 
(\ref{MCCmode}) and (\ref{MCCphi}). Precisely this term is supplied by the
part of the self-mass coming from the product of two 3-point vertices 
\cite{Glavan:2020gal}, but that is appealing to the exact calculation, rather
than explaining it stochastically or through the renormalization group. The
true explanation come through combining stochastic and RG effects, and it
may well involve following the running of the conformal coupling. 

\subsubsection{Einstein-Dirac}

Renormalizing single graviton loop corrections to the fermion self-energy 
requires three counterterms (\ref{DeltaDiracdS}),
\begin{eqnarray}
\lefteqn{\Delta \mathcal{L}_{\rm Dirac} \longrightarrow \alpha_1 
\overline{\widetilde{\psi}} (i \gamma^{\mu} \partial_{\mu} a^{-1})^3 a 
\widetilde{\psi} + \alpha_2 D (D\!-\!1) H^2 \overline{\widetilde{\psi}} i 
\gamma^{\mu} \partial_{\mu} \widetilde{\psi} } \nonumber \\
& & \hspace{7 cm} + \alpha_3 D (D \!-\! 1) H^2 \overline{\widetilde{\psi}} 
i \gamma^{k} \partial_{k} \widetilde{\psi} \; . \qquad \label{newDeltaDiracdS}
\end{eqnarray}
These are expressed on de Sitter background in terms of the conformally 
transformed fermion field $\widetilde{\psi} \equiv \psi/a^{(\frac{D-1}{2})}$.
The coefficients are \cite{Miao:2005am},
\begin{equation}
\alpha_1 = \tfrac{\kappa^2}{2^5 \pi^2} \tfrac{\mu^{D-4}}{D-4} \quad , \quad 
\alpha_2 R = -\tfrac{15 \kappa^2 H^2}{2^6 \pi^2} \tfrac{\mu^{D-4}}{D-4} 
\quad , \quad \alpha_3 R = \tfrac{7 \kappa^2 H^2}{2^5 \pi^2} \tfrac{\mu^{D-4}}{D-4} 
\; . \label{Diraccoefs}
\end{equation}
The term proportional to $\alpha_1$ actually includes a lower derivative
term because,
\begin{equation}
(\gamma^{\mu} \partial_{\mu} a^{-1})^2 = -\tfrac1{a^2} [\partial^2 + a H 
\gamma^0 \gamma^{\mu} \partial_{\mu} + 2 a H \partial_0] + H^2 \; .
\label{dslashsq}
\end{equation}
It follows that a curvature-dependent field strength renormalization is,
\begin{equation}
\delta Z = -\alpha_1 H^2 + \alpha_2 R \qquad \Longrightarrow \qquad 
\gamma = \tfrac{\partial \ln(1 + \delta Z)}{\partial \ln(\mu^2)} = -
\tfrac{17 \kappa^2 H^2}{128 \pi^2} \; . \label{gammaDirac}
\end{equation}
This would explain the graviton loop correction (\ref{GRDirac}) to the
fermion mode function if the latter could be considered a 1-point Green's
function, and the full resummation would be \cite{Miao:2025},
\begin{equation}
\widetilde{\psi}(x) \longrightarrow \widetilde{\psi}^{\rm tree} \times
[ \ln(a)]^{\frac{17 \kappa^2 H^2}{128 \pi^2}} \; . \label{fullpsi}
\end{equation}

The Dirac equation involves differentiated gravitons at order $\kappa^2$
through the spin connection,
\begin{eqnarray}
\lefteqn{-\tfrac12 \widetilde{e}^{\mu}_{~a} \gamma^a \widetilde{A}_{\mu bc} 
J^{bc} \widetilde{\psi} \sqrt{-\widetilde{g}} } \nonumber \\
& & \hspace{2cm} \longrightarrow \tfrac{\kappa^2}{4} \Bigl[ h 
h_{\mu \rho , \sigma} \!+\! \tfrac12 h^{\nu}_{~\rho} h_{\nu\sigma , \mu}
\!+\! (h^{\nu}_{~ \mu} h_{\nu\rho})_{,\sigma} \!+\! h^{\nu}_{~\sigma} 
h_{\mu \rho , \nu} \Bigr] \gamma^{\mu} J^{\rho\sigma} \widetilde{\psi} . 
\qquad \label{dgravpsi}
\end{eqnarray}
However, integrating these fields out gives zero \cite{Miao:2005am}. Hence 
there does not seem to be any stochastic contribution at 1-loop order.

\subsubsection{Einstein-Maxwell}

Recall the three counterterms needed to renormalize 1-loop graviton
contributions to the vacuum polarization \cite{Leonard:2013xsa},
\begin{eqnarray}
\lefteqn{\mathcal{L}_{\rm Maxwell} = C_4 D_{\alpha} F_{\mu\nu} D_{\beta} 
F_{\rho\sigma} g^{\alpha\beta} g^{\mu\rho} g^{\nu\sigma} \sqrt{-g}
+ \overline{C} H^2 F_{\mu\nu} F_{\rho\sigma} g^{\mu\rho} g^{\nu\sigma}
\sqrt{-g} } \nonumber \\
& & \hspace{7cm} + \Delta C H^2 F_{ij} F_{k\ell} g^{ik} g^{j\ell}
\sqrt{-g} \; . \qquad \label{newDeltaMaxwell}
\end{eqnarray} 
The coefficients are \cite{Leonard:2013xsa},
\begin{equation}
C_4 = \tfrac{\kappa^2}{96 \pi^2} \times \tfrac{\mu^{D-4}}{D-4} \quad , \quad
\overline{C} = \tfrac{7 \kappa^2}{96 \pi^2} \times \tfrac{\mu^{D-4}}{D-4} 
\quad , \quad \Delta C = -\tfrac{\kappa^2}{16 \pi^2} \times 
\tfrac{\mu^{D-4}}{D-4} \; . \label{Maxwellcoefs}
\end{equation}
By expanding out the covariant derivatives and specializing to de Sitter 
background the counterterms (\ref{newDeltaMaxwell}) can be rewritten as
\cite{Glavan:2023tet},
\begin{eqnarray}
\lefteqn{\mathcal{L}_{\rm Maxwell} = C_4 a^{D-6} \partial_{\alpha} F_{\mu\nu} 
\partial^{\alpha} F^{\mu\nu} + [\overline{C} - (3D \!-\! 8) C_4] a^{D-4} H^2
F_{\mu\nu} F^{\mu\nu} } \nonumber \\
& & \hspace{6.5cm} + [\Delta C - (D \!-\! 6) C_4] H^2 F_{ij} F_{k\ell} \; . 
\qquad \label{newerDeltaMaxwell}
\end{eqnarray}
The effective field strength renormalization is,
\begin{equation}
\delta Z = -4[\overline{C} - (3 D \!-\! 8) C_4] H^2 \qquad \Longrightarrow 
\qquad \gamma = \tfrac{\partial \ln(1 + \delta Z)}{\partial \ln(\mu^2)} =
-\tfrac{\kappa^2 H^2}{16 \pi^2} \; . \label{gammaMaxwell}
\end{equation}
If the electric field strength of radiation and the Coulomb potential are 
regarded as 2-point Green's functions and $\ln(a)$ is substituted for $\ln(\mu)$,
it can be seen that the Callan-Symanzik equation explains both (\ref{GRElectric}) 
and (\ref{GRCoulomb}). The full resummation is,
\begin{eqnarray}
F_{0i} &\!\!\! \longrightarrow \!\!\!& F^{\rm tree}_{0i} \times [a(t)]^{
\frac{\kappa^2 H^2}{8 \pi^2}} \; , \qquad \label{fullElectric} \\
\Phi &\!\!\! \longrightarrow \!\!\!& \tfrac{Q}{4\pi a r} \times 
[a(t) H r]^{\frac{\kappa^2 H^2}{8 \pi^2}} \; . \qquad \label{fullCoulomb}
\end{eqnarray}

\subsection{Graviton Corrections to Gravity}

The flat space limit of the simplest gauge \cite{Tsamis:1992xa,Woodard:2004ut}
implies that renormalizing the 1-loop graviton contribution to the graviton
self-energy must have at least the same two invariant counterterms 
(\ref{DeltaEinstein}) as matter loop contributions,
\begin{equation}
\Delta \mathcal{L}_{\rm Einstein} = c_1 R^2 \sqrt{-g} + c_2 
C^{\alpha\beta\gamma\delta} C_{\alpha\beta\gamma\delta} \sqrt{-g} \; .
\label{newDeltaEinstein}
\end{equation}
Because the coefficients are background-independent, they can be inferred
from the flat space limit \cite{Capper:1979ej,Tan:2021ibs},
\begin{equation}
c_1 = \tfrac{13}{2^6 \cdot 3^2 \cdot \pi^2} \times \tfrac{\mu^{D-4}}{D-4} 
\qquad , \qquad c_2 = \tfrac{61}{2^6 \cdot 3 \cdot 5 \cdot \pi^2} \times
\tfrac{\mu^{D-4}}{D-4} \; . \label{GRcoeffs}
\end{equation}
There will almost certainly be noncovariant counterterms arising from the 
de Sitter breaking gauge, the same way there were for the various graviton 
loop contributions to matter (\ref{newDeltaMMC}), (\ref{newDeltaMCCconf}), 
(\ref{newDeltaDiracdS}) and (\ref{newDeltaMaxwell}). However, the 
effective field strength renormalization from just the invariant terms
(\ref{newDeltaEinstein}) would be,
\begin{equation}
\delta Z = (24 c_1 - 4 c_2) \kappa^2 H^2 \qquad \Longrightarrow \qquad 
\gamma = \tfrac{\partial \ln(1 + \delta Z)}{\partial \ln(\mu^2)} = 
\tfrac{23 \kappa^2 H^2}{160 \pi^2} \; . \label{gammaGR}
\end{equation}

Note that the 1-loop finiteness of the flat space S-matrix for pure gravity 
\cite{tHooft:1974toh} in no way precludes field strength renormalizations 
of the sort (\ref{gammaGR}). Indeed, the flat space S-matrix of the nonlinear
sigma model (\ref{LAB}) also vanishes \cite{Woodard:2023rqo}, yet this model
still requires a curvature-dependent field strength renormalization whose
associated gamma function is (\ref{gammaB}). And section 4.1 describes how
the Callan-Symanzik equation (\ref{CSeqn}) not only explains the 1-loop
correction to the exchange potential (\ref{Bpotential}) but also implies
the resummation (\ref{fullBpot}).

It should also be noted that these renormalization groups effects can induce
at most a single large logarithm at 1-loop order. If the double logarithm 
corrections to the mode function (\ref{GRmode}) are correct then stochastic 
effects completely dominate those from the renormalization group. The same 
comments pertain for the triple logarithm corrections to the Newtonian 
potential (\ref{GRNewton}).

\section{The Gauge Issue}

The numerous matter loop corrections to gravity reported in expressions 
(\ref{MMCWeyl}-\ref{MMCSlip}), (\ref{MCCWeyl}-\ref{MCCSlip}), 
(\ref{DiracWeyl}-\ref{DiracSlip}) and (\ref{EMWeyl}-\ref{EMSlip}) involve
no graviton propagators and are completely gauge independent. However, 
loop corrections which involve graviton propagators do depend on the 
scheme by which the graviton gauge is fixed and this has led to a debate
about the reality of the large logarithms these corrections induce
\cite{Unruh:1998ic,Abramo:2001dc,Geshnizjani:2002wp,Garriga:2007zk,
Tsamis:2007is,Higuchi:2011vw,Miao:2011ng,Tanaka:2012wi,Tanaka:2013xe,
Tanaka:2013caa,Tanaka:2014ina}. It is so difficult to compute
graviton loops on de Sitter that almost all calculations have been made
using the simplest gauge \cite{Tsamis:1992xa,Woodard:2004ut} which was
reviewed in section 2.2. The single exception was of the vacuum 
polarization in the 1-parameter family of de Sitter invariant gauges 
\cite{Glavan:2015ura}. When the result was used to quantum-correct 
Maxwell's equation (\ref{Maxwell}) the result for 1-loop corrections to 
electromagnetic radiation had the same sign and functional form as 
(\ref{GRElectric}), with no dependence on the de Sitter invariant gauge 
parameter, but with a different numerical coefficient from the de Sitter
breaking gauge \cite{Glavan:2016bvp}. It therefore seems likely that 
large logarithmic corrections are real, but that a gauge independent 
technique must be employed to derive reliable results for their 
numerical coefficients.

This section is devoted to explaining how gauge dependence can be 
removed from the effective field equations by accounting for the quantum
gravitational correlations of the source which excites the effective 
field and the observer who detects it \cite{Miao:2017feh}.\footnote{One
can also remove gauge dependence by employing Green's functions based on
the expectation values of invariant operators \cite{Tsamis:1989yu,
Modanese:1994wv,Rovelli:2001my,Giddings:2005id,Giddings:2007nu,
Green:2008kj,Khavkine:2011kj,Donnelly:2015hta,Marolf:2015jha,Frob:2017apy,
Frob:2017gyj,Becker:2018quq,Becker:2019tlf}.} The procedure is to 
build the same diagrams that would go into an S-matrix element, and 
then simplify them with a series of relations derived by Donoghue 
\cite{Donoghue:1993eb,Donoghue:1994dn,Donoghue:1996mt} to capture the 
infrared physics. In the end, few traces of the source and observer remain, 
and each of the simplified diagrams can be regarded as a correction to the 
1PI 2-point function in the linearized, effective field equation. The 
section first describes how this procedure works on flat space, then 
discusses the generalization to de Sitter background. The section closes 
by reviewing what remains to be done.

\subsection{\normalsize How It Works on Flat Space}

Consider graviton corrections to a massless, minimally coupled scalar 
on flat space background ($g_{\mu\nu} \equiv \eta_{\mu\nu} + \kappa
h_{\mu\nu}$) in the 2-parameter family of covariant gauge fixing functions,
\begin{equation}
\mathcal{L}_{GF} = -\tfrac1{2 \alpha} \eta^{\mu\nu} F_{\mu} F_{\nu} \qquad , 
\qquad F_{\mu} = \eta^{\rho\sigma} \Bigl( h_{\mu \rho , \sigma} - 
\tfrac{\beta}{2} h_{\rho\sigma , \mu} \Bigr) \; . \label{gauge}
\end{equation}
In the Schwinger-Keldysh formalism the renormalized self-mass is 
\cite{Miao:2017feh},
\begin{eqnarray}
M_0^2(x;x') & = & -\mathcal{C}_0(\alpha,\beta) \times 
\tfrac{\kappa^2 \partial^8}{128 \pi^3} \Bigl\{ \theta(\Delta t \!-\! 
\Delta r) \Bigl( \ln[\mu^2 (\Delta t^2 \!-\! \Delta r^2)] - 1\Bigr) 
\Bigr\} , \label{M0} \qquad \\
\mathcal{C}_0(\alpha,\beta) & = & +\tfrac34 -\tfrac34 \times \alpha - 
\tfrac32 \times \tfrac1{\beta - 2} + \tfrac34 \times 
\tfrac{(\alpha - 3)}{(\beta - 2)^2} \; , \label{C0}
\end{eqnarray}
where $\Delta t \equiv t - t'$ and $\Delta r \equiv \Vert \vec{x} - 
\vec{x}' \Vert$. The linearized effective field equation,
\begin{equation}
\partial^2 \Phi(x) - \int \!\! d^4x \, M_0^2(x;x') \Phi(x') = J(x) \; ,
\end{equation}
is real and causal, but highly gauge-dependent.
\vskip 1cm
\begin{figure}[ht]
\includegraphics[width=4.0cm]{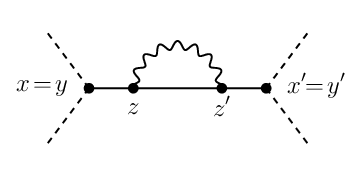} \hspace{0.5cm}
\includegraphics[width=4cm]{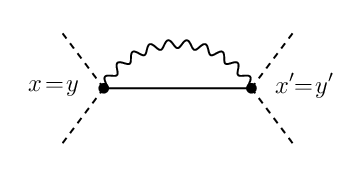} \hspace{0.5cm}
\includegraphics[width=4cm]{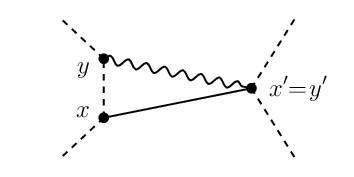}
\caption{\footnotesize Diagrams which contribute to graviton 
(wavy lines) loop corrections to the exchange of a massless scalar 
(solid lines) between two massive scalars (dashed lines). The left
hand diagram is $i=0$, the center diagram is $i=1$, and the right 
hand diagram is $i=2$ (which has four permutations).}
\label{Diagrams012}
\end{figure}

\begin{figure}[ht]
\includegraphics[width=4cm]{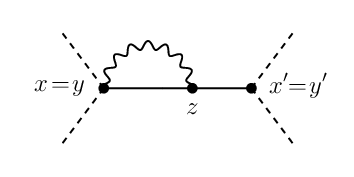} \hspace{0.5cm}
\includegraphics[width=4cm]{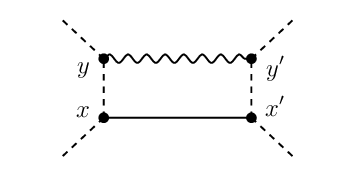} \hspace{0.5cm}
\includegraphics[width=4cm]{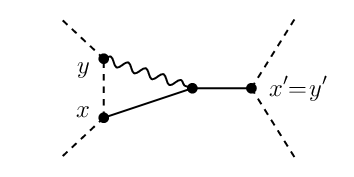}
\caption{\footnotesize Diagrams which contribute to graviton 
(wavy lines) loop corrections to the exchange of a massless scalar 
(solid lines) between two massive scalars (dashed lines). The left
hand diagram is $i=3$ (two permutations), the center diagram is 
$i=4$ (four permutations), and the right hand diagram is $i=5$ 
(four permutations).}
\label{Diagrams345}
\end{figure}

Now imagine quantum gravitational corrections to the scattering of 
two massive scalars ($\psi$) by the exchange of such a massless scalar
($\phi$) whose Lagrangian is,
\begin{equation}
\mathcal{L} = -\tfrac12 \partial_{\mu} \phi \partial_{\nu} \phi 
g^{\mu\nu} \sqrt{-g} - \tfrac12 \partial_{\mu} \psi \partial_{\nu}
\psi g^{\mu\nu} \sqrt{-g} - \tfrac12 m^2 \psi^2 \sqrt{-g} - \tfrac12
\lambda \phi \psi^2 \sqrt{-g} \; . \label{phipsi}
\end{equation}
Figures~\ref{Diagrams012} and \ref{Diagrams345} show the six classes of
diagrams which contribute. In position space these diagrams consist of 
products of (possibly differentiated) massive and massless propagators,
$i\Delta_m(x;x')$ and $i\Delta(x;x')$, respectively. For example, the
final diagram of Figure~\ref{Diagrams012} is,
\begin{eqnarray}
\lefteqn{ -i\kappa \Bigl[ -\overline{\partial}^{\mu}_{y} \partial^{\nu}_{y}
+ \tfrac12 \eta^{\mu\nu} (\overline{\partial}_y \!\cdot\! \partial_y + m^2)
\Bigr] i\Delta_m(y;x) } \nonumber \\
& & \hspace{4cm} \times -i\lambda i\Delta(x;x') \!\times\!
i [\mbox{}_{\mu\nu} \Delta_{\rho\sigma}](y;x') \!\times\! -\tfrac{i}{2}
\kappa \lambda \eta^{\rho\sigma} \; . \qquad \label{Diagram2}
\end{eqnarray}
Note that a barred derivative indicates that it is acted on the external 
state wave function. Also note that the graviton propagator $i 
[\mbox{}_{\mu\nu} \Delta_{\rho\sigma}](y;x')$ can be expressed in terms
of the massless scalar propagator $i\Delta(y;x')$. For example, with 
$\alpha = \beta = 1$ it is,
\begin{equation}
i [\mbox{}_{\mu\nu} \Delta_{\rho\sigma}](y;x') = \Bigl[ 2 \eta_{\mu (\rho}
\eta_{\sigma) \nu} - \tfrac{2}{D-2} \eta_{\mu\nu} \eta_{\rho\sigma}\Bigr]
i\Delta(y;x') \; . \label{simpleprop}
\end{equation}

All the 3-point and 4-point diagrams can be reduced to 2-point form by 
applying the Donoghue Identities \cite{Miao:2017feh,Glavan:2024elz}. The
one relevant to 3-point diagrams such as (\ref{Diagram2}) is,
\begin{equation}
i\Delta_m(x;y) i\Delta(x;x') i\Delta(y;x') \longrightarrow
\tfrac{i\delta^D(x - y)}{2 m^2} \, [ i\Delta(x;x')]^2 \; . \label{IntID1}
\end{equation}
Any 2-point contribution so obtained can be regarded as a correction 
to the self-mass through an identity based the massless propagator 
equation $\partial^2 i\Delta(x;x') = i\delta^D(x - x')$,
\begin{equation}
-i\lambda^2 f(x;x') = (-i \lambda)^2 \int d^Dz \, i\Delta(x;z) \int d^Dz' 
\, i\Delta(x';z') \times -i \partial_z^2 \partial_{z'}^2 f(z;z') \; . 
\label{keytrick}
\end{equation}
Because (\ref{keytrick}) has the same topology as the contribution from
$-i M^2_0(x;x')$ shown on the left hand side of Figure~\ref{Diagrams012}, 
one can consider $-i \partial_x^2 \partial_{x'}^2 f(x;x')$ as a correction
to the self-mass. By Poincar\'e invariance this correction takes the same 
form as (\ref{M0}), but with a different gauge-dependent coefficient, 
$\mathcal{C}_i(\alpha,\beta)$ which can be expressed in the same basis as
(\ref{C0}).

Table~\ref{Cab} collects the coefficients $\mathcal{C}_i(\alpha,\beta)$ 
for the six diagrams (plus permutations) depicted in Figures~\ref{Diagrams012} 
and \ref{Diagrams345}. Each contribution is labeled by the same number $i$ 
that was used in the figures, and each is given a description 
explaining the sort of quantum gravitational correlation it represents. For 
example, the description ``vertex-vertex'' for $i=1$ (the middle diagram of 
Figure~\ref{Diagrams012}) indicates that this contribution derives from
quantum gravitational correlations between the two $-i\lambda \phi \psi^2 
\sqrt{-g}$ vertices. Note especially that all dependence on the gauge 
parameters $\alpha$ and $\beta$ cancels out in the total.
\begin{table}[H]
\setlength{\tabcolsep}{8pt}
\def\arraystretch{1.5}
\centering
\begin{tabular}{|@{\hskip 1mm }c@{\hskip 1mm }||c|c|c|c|c|}
\hline
$i$ & $1$ & $\alpha$ & $\frac1{\beta-2}$ & $\frac{(\alpha-3)}{(\beta-2)^2}$ &
{\rm Description} \\
\hline\hline
0 & $+\frac34$ & $-\frac34$ & $-\frac32$ & $+\frac34$ &
{\rm scalar\ exchange} \\
\hline
1 & $0$ & $0$ & $0$ & $+1$ & {\rm vertex-vertex} \\
\hline
2 & $0$ & $0$ & $0$ & $0$ & {\rm vertex-source,observer} \\
\hline
3 & $0$ & $0$ & $+3$ & $-2$ & {\rm vertex-scalar} \\
\hline
4 & $+\frac{17}4$ & $-\frac34$ & $0$ & $-\frac14$ &
{\rm source-observer} \\
\hline
5 & $-2$ & $+\frac32$ & $-\frac32$ & $+\frac12$ &
{\rm scalar-source,observer} \\
\hline\hline
Total & $+3$ & $0$ & $0$ & $0$ & \\
\hline
\end{tabular}
\caption{\footnotesize The gauge dependent factors $C_i(\alpha,\beta)$ 
for each contribution to the invariant scalar self-mass-squared.}
\label{Cab}
\end{table}

The same procedure can be used to remove gauge dependence from the
linarized effective field equation of any massless field. The steps are: 
\begin{enumerate*}
\item{Write down (in position space) the diagrams which contribute to
quantum gravitational corrections to the scattering of two massive 
particles through the exchange of the massless field in question;}
\item{Reduce the 3-point and 4-point diagrams to 2-point form using
the Donoghue Identities; and}
\item{Regard each 2-point contribution as a correction to the 1PI 2-point
function of the massless field by the relation (\ref{keytrick}).}
\end{enumerate*} 
Applying these steps to electromagnetism results in a real, causal and gauge
independent modification of Maxwell's equation \cite{Katuwal:2021thy},
\begin{equation}
\partial_{\nu} F^{\nu\mu}(x) \!+\! \tfrac{5 \kappa^2 \partial^6}{768 \pi^3} 
\!\! \int \!\! d^4x' \, \theta(\Delta t \!-\! \Delta r) \Bigl( \ln[\mu^2
(\Delta t^2 \!-\! \Delta r^2)] \!-\! 1 \Bigr) \partial'_{\nu} 
F^{\nu\mu} = J^{\mu} . \label{QMax} 
\end{equation}
This equation can be solved the same as in classical electromagnetism.
For example, taking $J^{\mu}(t,\vec{x}) = Q \delta^{\mu}_{0} 
\delta^3(\vec{x})$ results in the same quantum gravitational correction 
to the Coulomb potential that one finds from computing the S-matrix and
then reconstructing the potential by inverse scattering 
\cite{Bjerrum-Bohr:2002aqa}.

\subsection{\normalsize How It Works on de Sitter}

It would be hugely surprising if flat space were the only background for 
which gauge dependence could be removed. Indeed, step 1 can be carried 
out for any geometry; it is not necessary either that the S-matrix is
observable or that it even exists. Step 3 can similarly be implemented
on any background, with the appropriate propagator equation of course. 
The spacetime dependence of the 1PI 2-point function is not as tightly
constrained on a general background, so that the various 2-point 
contributions will not be proportional to one another, but that was not
essential. 

The challenge is extending the Donoghue Identities to a general background. 
Donoghue derived there relations by computing the flat space amplitude and 
then extracting the part nonanalytic in the $t$ channel exchange momentum 
\cite{Donoghue:1993eb,Donoghue:1994dn,Donoghue:1996mt}. That cannot be done 
for cosmology but general coordinate invariance and the flat space 
correspondence limit may suffice. For example, the massive and massless 
propagators are scalars for any background, so the 3-point relation 
(\ref{IntID1}) generalizes to,
\begin{equation}
i\Delta_{m}(x;y) i\Delta(x;x') i\Delta(y;x') \longrightarrow 
\tfrac{i \delta^D(x - y)}{2 m^2 \sqrt{-g(x)}} \, [i \Delta(x;x')]^2 \; .
\label{newID1}
\end{equation}

Section 2.3.1 of this article reports on the single graviton correction to
the massless, minimally coupled scalar self-mass in the simplest gauge
\cite{Tsamis:1992xa,Woodard:2004ut}. Because the scalar exchange potential
(\ref{latepot}) shows a large logarithm \cite{Glavan:2021adm}, this
system was identified as a good venue for the first implementation of the 
3-step procedure on de Sitter background. The simplest gauge was employed 
to compute the same 5 diagrams which appear in Table~\ref{Cab}, plus some
diagrams involving coincident massless propagators, which vanish in flat
space). The Donoghue Identities were minimally extended based on general 
coordinate invariance. Although the numerical coefficient of the large 
logarithm in the 1-loop exchange potential does change, it is still nonzero, 
which strongly supports the reality of graviton-induced logarithms 
\cite{Glavan:2024elz}.

\subsection{\normalsize What Remains to Do}

There can be little doubt that this technique is correct on flat space
background owing to the close connection between it and the S-matrix, 
and to the explicit cancellation of dependence on the parameters $\alpha$ 
and $\beta$ for a massless, minimally coupled scalar \cite{Miao:2017feh} 
and electromagnetism \cite{Katuwal:2021thy}. The only problematic point
in generalizing it to cosmology is the Donoghue Identities. For this 
reason, it is necessary to repeat the de Sitter calculation, which was 
already done in the simplest gauge \cite{Glavan:2024elz}, in a general
2-parameter family of gauges analogous to (\ref{gauge}). The order
$\alpha$ and $\beta$ corrections to the graviton propagator have been 
derived \cite{Glavan:2019msf}, and work is far advanced on using them to
construct the analog of Table~\ref{Cab}. If the same cancellation of 
dependence on $\alpha$ and $\beta$ occurs, the validity of the technique 
can be regarded as well established.

Once the technique's validity has been demonstrated, it must be used to
derive gauge-independent results for all the known large graviton loop
corrections reviewed in section 2.3. To facilitate this massive 
undertaking, it is important to simplify the computational procedure 
as much as possible. Extensive simplifications were discovered during the
recent computation on de Sitter background \cite{Glavan:2024elz}. For 
example, in any gauge and on any background, parts of the $i=4$ 
contributions cancel those of $i=1$ and $i=2$. Similar cancellations 
occur between $i=3$ and $i=5$. It may even be possible to view a change 
of gauge in the $i=0$ contribution as automatically being canceled by 
combinations of the higher diagrams, without performing any explicit 
computations.

\section{Conclusions}

The first graviton-induced logarithm from a dimensionally regulated and fully
renormalized computation was discovered just 20 years ago \cite{Miao:2005am,
Miao:2006gj}. Since then many others have been found, in graviton corrections 
to matter \cite{Kahya:2007bc,Leonard:2013xsa,Glavan:2013jca,Wang:2014tza,
Glavan:2020gal,Glavan:2020ccz,Glavan:2021adm} and to gravity \cite{Tan:2021ibs,
Tan:2021lza,Tan:2022xpn}. The reciprocal process of matter corrections to 
gravity also produces large logarithms \cite{Wang:2015eaa,Miao:2024atw,
Foraci:2024vng,Foraci:2024cwi}. These effects were reviewed in section 2.3.

A fascinating aspect of these corrections is that they grow, often in time and
sometimes in space. During a prolonged period of inflation this growth must
eventually cause perturbation theory to break down. Learning what happens next
requires a nonperturbative resummation scheme. The search for such a technique
has been long and confusing because the large logarithms derive from two 
distinct sources, the tail part of the graviton propagator (\ref{4Dtail}) and
the incomplete cancellation between primitive divergences and counterterms
(\ref{RGsource}). Resumming the logarithms from each source requires its own
technique. That fact was not recognized until 2018 \cite{Miao:2018bol};
before then every attempt to apply the resummation schemes adapted to one 
source failed when confronted with effects induced by the other source
\cite{Miao:2008sp,Woodard:2008yt}. It was not until 2021 that a series of 
1-loop and 2-loop computations involving nonlinear sigma models on de Sitter
background provided a sufficiently rich and diverse source of effects to see
both sources at work, and to devise appropriate resummation techniques
\cite{Miao:2021gic}.

Section 3 explains how logarithms induced by the tail (\ref{4Dtail}) can be 
resummed using a variant of Starobinsky's stochastic formalism 
\cite{Starobinsky:1986fx,Starobinsky:1994bd}. The variation is that one must
first integrate out passive fields (which have no tail) and differentiated 
active fields (which do have a tail) in the presence of a constant active
field background \cite{Miao:2006pn,Prokopec:2007ak,Miao:2021gic,Miao:2024nsz,
Miao:2024shs}. The result is a scalar potential model of the form appropriate
to be treated by Starobinsky's formalism. A noteworthy technical advance 
associated with this work is that integrating out differentiated active 
fields induces novel effective potentials which derive from changes in field 
strengths \cite{Miao:2021gic} and changes in the background geometry 
\cite{Miao:2024nsz,Miao:2024shs}, rather than from changes in the mass.
 
Section 4 explains how logarithms from the second source (\ref{RGsource}) can 
be resummed using a variant of the renormalization group. The variation is 
that the BPHZ counterterms needed to renormalize 1PI 2-point functions contain
a part which can be viewed as a curvature-dependent renormalization of bare
parameters \cite{Miao:2021gic,Glavan:2021adm,Glavan:2023tet,Miao:2024nsz,
Foraci:2024vng,Foraci:2024cwi}. One can then explain, and sometimes resum, the
large logarithms using the Callan-Symanzik equation, with the usual replacement 
of derivatives with respect to logarithms of the renormalization scale by 
derivatives with respect to logarithms of dynamical quantities such as the 
scale factor $a(t)$ and the co-moving coordinate distance $H r$.

Matter loop corrections to gravity are completely independent of the 
graviton gauge, and the fact that they induce the same sorts of large 
logarithms as are found from graviton loop corrections in a fixed gauge, 
is strong evidence that graviton-induced logarithms are real. Indeed, 
one might suspect that this is required by the reciprocity between action 
and reaction. However, it is necessary to derive reliable predictions from 
graviton loop corrections and section 5 describes a procedure for doing so 
by accounting for the quantum gravitational correlations with the source 
which excites the effective field and the observer who measures it 
\cite{Miao:2017feh}. The technique is to form the same set of 
position-space amplitudes which would contribute to the $t$ channel exchange
of the light field between two massive particles. This amounts to a set 
of 2-point, 3-point and 4-point diagrams such as those pictured in 
Figures~\ref{Diagrams012} and \ref{Diagrams345}. The 3-point and 4-point 
diagrams can be reduced to 2-point form using identities derived by
Donoghue and collaborators to extract the parts not analytic in the exchange
momentum \cite{Donoghue:1993eb,Donoghue:1994dn,Donoghue:1996mt}. Finally, 
the various 2-point diagrams are expressed in a form (\ref{keytrick}) which
can be regarded as correcting the gauge-dependent 1PI 2-point function. 
When this technique is implemented on flat space background one can 
explicitly demonstrate the cancellation of gauge dependence for graviton
corrections to a massless, minimally coupled scalar \cite{Miao:2017feh} and
to electromagnetism \cite{Katuwal:2021thy}. It has been implemented, in
the simplest gauge, on de Sitter background, for the massless, minimally
coupled scalar \cite{Glavan:2024elz} and the result seems to confirm the
reality of graviton-induced logarithms. Work is far advanced on an explicit
demonstration of gauge independence in a 2-parameter family of gauges
\cite{Glavan:2019msf}.

Despite the impressive progress in this field, much remains to be done.
One issue is mixing between stochastic and renormalization group effects
in models which show both. The obvious guess is that one should employ the
renormalization group to improve the stochastic field equations and then
use those. This needs to be checked in the simple context of nonlinear
sigma models.

Another area where more work is needed is simplifying the stochastic 
Langevin equation of pure gravity \cite{Miao:2024shs}. This was discussed
in section 3.4. These equations come from applying the 3-step procedure
given there to the four parts (\ref{ghost}), (\ref{L123}), (\ref{L45})
and (\ref{L6}) which make up the gauge-fixed Lagrangian. The first and 
second steps (varying the action and integrating out differentiated 
graviton fields) have been fully implemented \cite{Miao:2024shs}, but 
the $3+1$ decomposition associated with deriving Langevin kinetic operator
remains. One must also re-compute the 1PI 1-point and 2-point functions in
the new gauge (\ref{newgauge}) in order to confirm the consistency of the
formalism. 

More work is also required to sort out what is going on with the large
graviton-induced logarithms (\ref{MCCmode}) and (\ref{MCCphi}) of the
massless, conformally coupled scalar mode function and exchange potential.
As explained in section 4.2.2, these do not seem to be completely explained
using either the renromalization group or the stochastic formalism, 
although elements of both methods may pertain. It is also possible that
the explanation is a new sort of graviton effect in which the 
$\widetilde{\phi}^2 \widetilde{R} \sqrt{-\widetilde{g}}$ term of the 
Lagrangian (\ref{ConformalFields}) breaks the invariance under constant 
shifts of $\widetilde{\phi}$ which is present for de Sitter background.

One issue which does seem to be well in hand is the generalization of de
Sitter to realistic expansion histories $a(t)$. For the case of stochastic 
corrections one can use the approximations developed in \cite{Kasdagli:2023nzj} 
to evaluate the three key coincidence limits (\ref{2coinc}-\ref{3rdcoinc}).
For the case of renormalization group corrections, the graviton self-energy
is known for an arbitrary expansion history for contributions from massless, 
conformally coupled scalars and from massless Dirac fermions 
\cite{Foraci:2024cwi}. If one then writes the Lichnerowicz operator for a 
general cosmological background the resulting equations can be solved for a 
realistic cosmology, at least numerically.

\vskip .3cm

\centerline{\bf Acknowledgements}

It is a pleasure to acknowledge collaboration, conversation and 
correspondence on these subjects with A. J. Foraci, D. Glavan, S. P. Miao, 
T. Prokopec, N. C. Tsamis and B. Yesilyurt. This work was partially 
supported by NSF grant PHY-2207514 and by the Institute for Fundamental 
Theory at the University of Florida.


\begin{thebibliography}{99}

\bibitem{DES:2021wwk}
T.~M.~C.~Abbott \textit{et al.} [DES],
Phys. Rev. D \textbf{105}, no.2, 023520 (2022)
doi:10.1103/PhysRevD.105.023520
[arXiv:2105.13549 [astro-ph.CO]].

\bibitem{Kamionkowski:2022pkx}
M.~Kamionkowski and A.~G.~Riess,
Ann. Rev. Nucl. Part. Sci. \textbf{73}, 153-180 (2023)
doi:10.1146/annurev-nucl-111422-024107
[arXiv:2211.04492 [astro-ph.CO]].

\bibitem{Planck:2018vyg}
N.~Aghanim \textit{et al.} [Planck],
Astron. Astrophys. \textbf{641}, A6 (2020)
[erratum: Astron. Astrophys. \textbf{652}, C4 (2021)]
doi:10.1051/0004-6361/201833910
[arXiv:1807.06209 [astro-ph.CO]].

\bibitem{Geshnizjani:2011dk}
G.~Geshnizjani, W.~H.~Kinney and A.~Moradinezhad Dizgah,
JCAP \textbf{11}, 049 (2011)
doi:10.1088/1475-7516/2011/11/049
[arXiv:1107.1241 [astro-ph.CO]].

\bibitem{Kinney:2012zbd}
W.~H.~Kinney, G.~Geshnizjani and A.~Moradinezhad Dizgah,

\bibitem{Mukhanov:1981xt}
V.~F.~Mukhanov and G.~V.~Chibisov,
JETP Lett. \textbf{33}, 532-535 (1981)

\bibitem{Starobinsky:1979ty}
A.~A.~Starobinsky,
JETP Lett. \textbf{30}, 682-685 (1979)

\bibitem{Donoghue:1993eb}
J.~F.~Donoghue,
Phys. Rev. Lett. \textbf{72}, 2996-2999 (1994)
doi:10.1103/PhysRevLett.72.2996
[arXiv:gr-qc/9310024 [gr-qc]].

\bibitem{Donoghue:1994dn}
J.~F.~Donoghue,
Phys. Rev. D \textbf{50}, 3874-3888 (1994)
doi:10.1103/PhysRevD.50.3874
[arXiv:gr-qc/9405057 [gr-qc]].

\bibitem{Donoghue:2017ovt}
J.~Donoghue,
Scholarpedia \textbf{12}, no.4, 32997 (2017)
doi:10.4249/scholarpedia.32997

\bibitem{Miao:2006gj}
S.~P.~Miao and R.~P.~Woodard,
Phys. Rev. D \textbf{74}, 024021 (2006)
doi:10.1103/PhysRevD.74.024021
[arXiv:gr-qc/0603135 [gr-qc]].

\bibitem{Glavan:2013jca}
D.~Glavan, S.~P.~Miao, T.~Prokopec and R.~P.~Woodard,
Class. Quant. Grav. \textbf{31}, 175002 (2014)
doi:10.1088/0264-9381/31/17/175002
[arXiv:1308.3453 [gr-qc]].

\bibitem{Wang:2014tza}
C.~L.~Wang and R.~P.~Woodard,
Phys. Rev. D \textbf{91}, no.12, 124054 (2015)
doi:10.1103/PhysRevD.91.124054
[arXiv:1408.1448 [gr-qc]].

\bibitem{Glavan:2021adm}
D.~Glavan, S.~P.~Miao, T.~Prokopec and R.~P.~Woodard,
JHEP \textbf{03}, 088 (2022)
doi:10.1007/JHEP03(2022)088
[arXiv:2112.00959 [gr-qc]].

\bibitem{Tan:2021lza}
L.~Tan, N.~C.~Tsamis and R.~P.~Woodard,
Phil. Trans. Roy. Soc. Lond. A \textbf{380}, 0187 (2021)
doi:10.1098/rsta.2021.0187
[arXiv:2107.13905 [gr-qc]].

\bibitem{Tan:2022xpn}
L.~Tan, N.~C.~Tsamis and R.~P.~Woodard,
Universe \textbf{8}, no.7, 376 (2022)
doi:10.3390/universe8070376
[arXiv:2206.11467 [gr-qc]].

\bibitem{Radkowski:1970}
A.~F.~Radkowski,
Ann. Phys. \textbf{56}, no. 2, 319-354 (1970,
doi.org/10.1016/0003-4916(70)90021-7.

\bibitem{tHooft:1972tcz}
G.~'t Hooft and M.~J.~G.~Veltman,
Nucl. Phys. B \textbf{44}, 189-213 (1972)
doi:10.1016/0550-3213(72)90279-9

\bibitem{Bollini:1972ui}
C.~G.~Bollini and J.~J.~Giambiagi,
Nuovo Cim. B \textbf{12}, 20-26 (1972)
doi:10.1007/BF02895558

\bibitem{Bogoliubov:1957gp}
N.~N.~Bogoliubov and O.~S.~Parasiuk,
Acta Math. \textbf{97}, 227-266 (1957)
doi:10.1007/BF02392399

\bibitem{Hepp:1966eg}
K.~Hepp,
Commun. Math. Phys. \textbf{2}, 301-326 (1966)
doi:10.1007/BF01773358

\bibitem{Zimmermann:1968mu}
W.~Zimmermann,
Commun. Math. Phys. \textbf{11}, 1-8 (1968)
doi:10.1007/BF01654298

\bibitem{Zimmermann:1969jj}
W.~Zimmermann,
Commun. Math. Phys. \textbf{15}, 208-234 (1969)
doi:10.1007/BF01645676

\bibitem{DeWitt:1960fc}
B.~S.~DeWitt and R.~W.~Brehme,
Annals Phys. \textbf{9}, 220-259 (1960)
doi:10.1016/0003-4916(60)90030-0

\bibitem{Lifshitz:1945du}
E.~Lifshitz,
J. Phys. (USSR) \textbf{10}, no.2, 116 (1946)
doi:10.1007/s10714-016-2165-8

\bibitem{Onemli:2002hr}
V.~K.~Onemli and R.~P.~Woodard,
Class. Quant. Grav. \textbf{19}, 4607 (2002)
doi:10.1088/0264-9381/19/17/311
[arXiv:gr-qc/0204065 [gr-qc]].

\bibitem{Onemli:2004mb}
V.~K.~Onemli and R.~P.~Woodard,
Phys. Rev. D \textbf{70}, 107301 (2004)
doi:10.1103/PhysRevD.70.107301
[arXiv:gr-qc/0406098 [gr-qc]].

\bibitem{Miao:2021gic}
S.~P.~Miao, N.~C.~Tsamis and R.~P.~Woodard,
JHEP \textbf{03}, 069 (2022)
doi:10.1007/JHEP03(2022)069
[arXiv:2110.08715 [gr-qc]].

\bibitem{Allen:1987tz}
B.~Allen and A.~Folacci,
Phys. Rev. D \textbf{35}, 3771 (1987)
doi:10.1103/PhysRevD.35.3771

\bibitem{Allen:1986ta}
B.~Allen,
Phys. Rev. D \textbf{34}, 3670 (1986)
doi:10.1103/PhysRevD.34.3670

\bibitem{Allen:1986tt}
B.~Allen and M.~Turyn,
Nucl. Phys. B \textbf{292}, 813 (1987)
doi:10.1016/0550-3213(87)90672-9

\bibitem{Hawking:2000ee}
S.~W.~Hawking, T.~Hertog and N.~Turok,
Phys. Rev. D \textbf{62}, 063502 (2000)
doi:10.1103/PhysRevD.62.063502
[arXiv:hep-th/0003016 [hep-th]].

\bibitem{Higuchi:2001uv}
A.~Higuchi and S.~S.~Kouris,
Class. Quant. Grav. \textbf{18}, 4317-4328 (2001)
doi:10.1088/0264-9381/18/20/311
[arXiv:gr-qc/0107036 [gr-qc]].

\bibitem{Higuchi:2002sc}
A.~Higuchi and R.~H.~Weeks,
Class. Quant. Grav. \textbf{20}, 3005-3022 (2003)
doi:10.1088/0264-9381/20/14/303
[arXiv:gr-qc/0212031 [gr-qc]].

\bibitem{Morrison:2013rqa}
I.~A.~Morrison,
[arXiv:1302.1860 [gr-qc]].

\bibitem{Antoniadis:1986sb}
I.~Antoniadis and E.~Mottola,
J. Math. Phys. \textbf{32}, 1037-1044 (1991)
doi:10.1063/1.529381

\bibitem{Folacci:1992xc}
A.~Folacci,
Phys. Rev. D \textbf{46}, 2553-2559 (1992)
doi:10.1103/PhysRevD.46.2553
[arXiv:0911.2064 [gr-qc]].

\bibitem{Folacci:1996dv}
A.~Folacci,
Phys. Rev. D \textbf{53}, 3108-3117 (1996)
doi:10.1103/PhysRevD.53.3108

\bibitem{Miao:2009hb}
S.~P.~Miao, N.~C.~Tsamis and R.~P.~Woodard,
J. Math. Phys. \textbf{50}, 122502 (2009)
doi:10.1063/1.3266179
[arXiv:0907.4930 [gr-qc]].

\bibitem{Kiritsis:1994ta}
E.~Kiritsis and C.~Kounnas,
Nucl. Phys. B \textbf{442}, 472-493 (1995)
doi:10.1016/0550-3213(95)00156-M
[arXiv:hep-th/9501020 [hep-th]].

\bibitem{Janssen:2008px}
T.~M.~Janssen, S.~P.~Miao, T.~Prokopec and R.~P.~Woodard,
Class. Quant. Grav. \textbf{25}, 245013 (2008)
doi:10.1088/0264-9381/25/24/245013
[arXiv:0808.2449 [gr-qc]].

\bibitem{Miao:2013isa}
S.~P.~Miao, P.~J.~Mora, N.~C.~Tsamis and R.~P.~Woodard,
Phys. Rev. D \textbf{89}, no.10, 104004 (2014)
doi:10.1103/PhysRevD.89.104004
[arXiv:1306.5410 [gr-qc]].

\bibitem{Miao:2011fc}
S.~P.~Miao, N.~C.~Tsamis and R.~P.~Woodard,
J. Math. Phys. \textbf{52}, 122301 (2011)
doi:10.1063/1.3664760
[arXiv:1106.0925 [gr-qc]].

\bibitem{Mora:2012zi}
P.~J.~Mora, N.~C.~Tsamis and R.~P.~Woodard,
J. Math. Phys. \textbf{53}, 122502 (2012)
doi:10.1063/1.4764882
[arXiv:1205.4468 [gr-qc]].

\bibitem{Tsamis:1992xa}
N.~C.~Tsamis and R.~P.~Woodard,
Commun. Math. Phys. \textbf{162}, 217-248 (1994)
doi:10.1007/BF02102015

\bibitem{Woodard:2004ut}
R.~P.~Woodard,
[arXiv:gr-qc/0408002 [gr-qc]].

\bibitem{Tsamis:1996qm}
N.~C.~Tsamis and R.~P.~Woodard,
Annals Phys. \textbf{253}, 1-54 (1997)
doi:10.1006/aphy.1997.5613
[arXiv:hep-ph/9602316 [hep-ph]].

\bibitem{Tsamis:1996qk}
N.~C.~Tsamis and R.~P.~Woodard,
Phys. Rev. D \textbf{54}, 2621-2639 (1996)
doi:10.1103/PhysRevD.54.2621
[arXiv:hep-ph/9602317 [hep-ph]].

\bibitem{Tsamis:2005je}
N.~C.~Tsamis and R.~P.~Woodard,
Annals Phys. \textbf{321}, 875-893 (2006)
doi:10.1016/j.aop.2005.08.004
[arXiv:gr-qc/0506056 [gr-qc]].

\bibitem{Miao:2005am}
S.~P.~Miao and R.~P.~Woodard,
Class. Quant. Grav. \textbf{23}, 1721-1762 (2006)
doi:10.1088/0264-9381/23/5/016
[arXiv:gr-qc/0511140 [gr-qc]].

\bibitem{Kahya:2007bc}
E.~O.~Kahya and R.~P.~Woodard,
Phys. Rev. D \textbf{76}, 124005 (2007)
doi:10.1103/PhysRevD.76.124005
[arXiv:0709.0536 [gr-qc]].

\bibitem{Miao:2012bj}
S.~P.~Miao,
Phys. Rev. D \textbf{86}, 104051 (2012)
doi:10.1103/PhysRevD.86.104051
[arXiv:1207.5241 [gr-qc]].

\bibitem{Leonard:2013xsa}
K.~E.~Leonard and R.~P.~Woodard,
Class. Quant. Grav. \textbf{31}, 015010 (2014)
doi:10.1088/0264-9381/31/1/015010
[arXiv:1304.7265 [gr-qc]].

\bibitem{Glavan:2015ura}
D.~Glavan, S.~P.~Miao, T.~Prokopec and R.~P.~Woodard,
Class. Quant. Grav. \textbf{32}, no.19, 195014 (2015)
doi:10.1088/0264-9381/32/19/195014
[arXiv:1504.00894 [gr-qc]].

\bibitem{Miao:2017vly}
S.~P.~Miao, N.~C.~Tsamis and R.~P.~Woodard,
Phys. Rev. D \textbf{95}, no.12, 125008 (2017)
doi:10.1103/PhysRevD.95.125008
[arXiv:1702.05694 [gr-qc]].

\bibitem{Glavan:2020gal}
D.~Glavan, S.~P.~Miao, T.~Prokopec and R.~P.~Woodard,
Phys. Rev. D \textbf{101}, no.10, 106016 (2020)
doi:10.1103/PhysRevD.101.106016
[arXiv:2003.02549 [gr-qc]].

\bibitem{Chou:1984es}
K.~c.~Chou, Z.~b.~Su, B.~l.~Hao and L.~Yu,
Phys. Rept. \textbf{118}, 1-131 (1985)
doi:10.1016/0370-1573(85)90136-X

\bibitem{Jordan:1986ug}
R.~D.~Jordan,
Phys. Rev. D \textbf{33}, 444-454 (1986)
doi:10.1103/PhysRevD.33.444

\bibitem{Calzetta:1986ey}
E.~Calzetta and B.~L.~Hu,
Phys. Rev. D \textbf{35}, 495 (1987)
doi:10.1103/PhysRevD.35.495

\bibitem{Ford:2004wc}
L.~H.~Ford and R.~P.~Woodard,
Class. Quant. Grav. \textbf{22}, 1637-1647 (2005)
doi:10.1088/0264-9381/22/9/011
[arXiv:gr-qc/0411003 [gr-qc]].

\bibitem{Glavan:2019yfc}
D.~Glavan, S.~P.~Miao, T.~Prokopec and R.~P.~Woodard,
Phys. Lett. B \textbf{798}, 134944 (2019)
doi:10.1016/j.physletb.2019.134944
[arXiv:1908.11113 [gr-qc]].

\bibitem{tHooft:1974toh}
G.~'t Hooft and M.~J.~G.~Veltman,
Ann. Inst. H. Poincare A Phys. Theor. \textbf{20}, 69-94 (1974)

\bibitem{Park:2011ww}
S.~Park and R.~P.~Woodard,
Phys. Rev. D \textbf{83}, 084049 (2011)
doi:10.1103/PhysRevD.83.084049
[arXiv:1101.5804 [gr-qc]].

\bibitem{Miao:2024atw}
S.~P.~Miao, N.~C.~Tsamis and R.~P.~Woodard,
JHEP \textbf{07}, 099 (2024)
doi:10.1007/JHEP07(2024)099
[arXiv:2405.00116 [gr-qc]].

\bibitem{Boran:2014xpa}
S.~Boran, E.~O.~Kahya and S.~Park,
Phys. Rev. D \textbf{90}, no.12, 124054 (2014)
doi:10.1103/PhysRevD.90.124054
[arXiv:1409.7753 [gr-qc]].

\bibitem{Boran:2017fsx}
S.~Boran, E.~O.~Kahya and S.~Park,
Phys. Rev. D \textbf{96}, no.2, 025001 (2017)
doi:10.1103/PhysRevD.96.025001
[arXiv:1704.05880 [gr-qc]].

\bibitem{Frob:2017apy}
M.~B.~Fr\"ob,
Class. Quant. Grav. \textbf{35}, no.3, 035005 (2018)
doi:10.1088/1361-6382/aa9ad1
[arXiv:1706.01891 [hep-th]].

\bibitem{Glavan:2020ccz}
D.~Glavan, S.~P.~Miao, T.~Prokopec and R.~P.~Woodard,
Phys. Rev. D \textbf{103}, no.10, 105022 (2021)
doi:10.1103/PhysRevD.103.105022
[arXiv:2007.10395 [gr-qc]].

\bibitem{Capper:1973bk}
D.~M.~Capper,
Nuovo Cim. A \textbf{25}, 29 (1975)
doi:10.1007/BF02735608

\bibitem{Foraci:2024cwi}
A.~J.~Foraci and R.~P.~Woodard,
[arXiv:2501.01972 [gr-qc]].

\bibitem{Woodard:1984sj}
R.~P.~Woodard,
Phys. Lett. B \textbf{148}, 440-444 (1984)
doi:10.1016/0370-2693(84)90734-2

\bibitem{Feinberg:1968zz}
G.~Feinberg and J.~Sucher,
Phys. Rev. \textbf{166}, 1638-1644 (1968)
doi:10.1103/PhysRev.166.1638

\bibitem{Hsu:1992tg}
S.~D.~H.~Hsu and P.~Sikivie,
Phys. Rev. D \textbf{49}, 4951-4953 (1994)
doi:10.1103/PhysRevD.49.4951
[arXiv:hep-ph/9211301 [hep-ph]].

\bibitem{Capper:1973mv}
D.~M.~Capper and M.~J.~Duff,
Nucl. Phys. B \textbf{82}, 147-154 (1974)
doi:10.1016/0550-3213(74)90582-3

\bibitem{Duff:2000mt}
M.~J.~Duff and J.~T.~Liu,
Phys. Rev. Lett. \textbf{85}, 2052-2055 (2000)
doi:10.1088/0264-9381/18/16/310
[arXiv:hep-th/0003237 [hep-th]].

\bibitem{Capper:1974ed}
D.~M.~Capper, M.~J.~Duff and L.~Halpern,
Phys. Rev. D \textbf{10}, 461-467 (1974)
doi:10.1103/PhysRevD.10.461

\bibitem{Wang:2015eaa}
C.~L.~Wang and R.~P.~Woodard,
Phys. Rev. D \textbf{92}, 084008 (2015)
doi:10.1103/PhysRevD.92.084008
[arXiv:1508.01564 [gr-qc]].

\bibitem{Foraci:2024vng}
A.~J.~Foraci and R.~P.~Woodard,
[arXiv:2412.11022 [gr-qc]].

\bibitem{Deser:1974cz}
S.~Deser and P.~van Nieuwenhuizen,
Phys. Rev. D \textbf{10}, 401 (1974)
doi:10.1103/PhysRevD.10.401

\bibitem{Deser:1974zzd}
S.~Deser and P.~van Nieuwenhuizen,
Phys. Rev. Lett. \textbf{32}, 245-247 (1974)
doi:10.1103/PhysRevLett.32.245

\bibitem{Tsamis:2023fri}
N.~C.~Tsamis, R.~P.~Woodard and B.~Yesilyurt,
Phys. Lett. B \textbf{849}, 138472 (2024)
doi:10.1016/j.physletb.2024.138472
[arXiv:2312.15913 [gr-qc]].

\bibitem{Tan:2021ibs}
L.~Tan, N.~C.~Tsamis and R.~P.~Woodard,
Class. Quant. Grav. \textbf{38}, no.14, 145024 (2021)
doi:10.1088/1361-6382/ac0233
[arXiv:2103.08547 [gr-qc]].

\bibitem{Starobinsky:1986fx}
A.~A.~Starobinsky,
Lect. Notes Phys. \textbf{246}, 107-126 (1986)
doi:10.1007/3-540-16452-9\_6

\bibitem{Starobinsky:1994bd}
A.~A.~Starobinsky and J.~Yokoyama,
Phys. Rev. D \textbf{50}, 6357-6368 (1994)
doi:10.1103/PhysRevD.50.6357
[arXiv:astro-ph/9407016 [astro-ph]].

\bibitem{Tsamis:2005hd}
N.~C.~Tsamis and R.~P.~Woodard,
Nucl. Phys. B \textbf{724}, 295-328 (2005)
doi:10.1016/j.nuclphysb.2005.06.031
[arXiv:gr-qc/0505115 [gr-qc]].

\bibitem{Miao:2006pn}
S.~P.~Miao and R.~P.~Woodard,
Phys. Rev. D \textbf{74}, 044019 (2006)
doi:10.1103/PhysRevD.74.044019
[arXiv:gr-qc/0602110 [gr-qc]].

\bibitem{Prokopec:2007ak}
T.~Prokopec, N.~C.~Tsamis and R.~P.~Woodard,
Annals Phys. \textbf{323}, 1324-1360 (2008)
doi:10.1016/j.aop.2007.08.008
[arXiv:0707.0847 [gr-qc]].

\bibitem{Candelas:1975du}
P.~Candelas and D.~J.~Raine,
Phys. Rev. D \textbf{12}, 965-974 (1975)
doi:10.1103/PhysRevD.12.965

\bibitem{Tsamis:2006gj}
N.~C.~Tsamis and R.~P.~Woodard,
J. Math. Phys. \textbf{48}, 052306 (2007)
doi:10.1063/1.2738361
[arXiv:gr-qc/0608069 [gr-qc]].

\bibitem{Prokopec:2006ue}
T.~Prokopec, N.~C.~Tsamis and R.~P.~Woodard,
Class. Quant. Grav. \textbf{24}, 201-230 (2007)
doi:10.1088/0264-9381/24/1/011
[arXiv:gr-qc/0607094 [gr-qc]].

\bibitem{Prokopec:2008gw}
T.~Prokopec, N.~C.~Tsamis and R.~P.~Woodard,
Phys. Rev. D \textbf{78}, 043523 (2008)
doi:10.1103/PhysRevD.78.043523
[arXiv:0802.3673 [gr-qc]].

\bibitem{Kitamoto:2010et}
H.~Kitamoto and Y.~Kitazawa,
Phys. Rev. D \textbf{83}, 104043 (2011)
doi:10.1103/PhysRevD.83.104043
[arXiv:1012.5930 [hep-th]].

\bibitem{Kitamoto:2011yx}
H.~Kitamoto and Y.~Kitazawa,
Phys. Rev. D \textbf{85}, 044062 (2012)
doi:10.1103/PhysRevD.85.044062
[arXiv:1109.4892 [hep-th]].

\bibitem{Kitamoto:2018dek}
H.~Kitamoto,
Phys. Rev. D \textbf{100}, no.2, 025020 (2019)
doi:10.1103/PhysRevD.100.025020
[arXiv:1811.01830 [hep-th]].

\bibitem{Woodard:2023rqo}
R.~P.~Woodard and B.~Yesilyurt,
JHEP \textbf{06}, 206 (2023)
doi:10.1007/JHEP06(2023)206
[arXiv:2302.11528 [gr-qc]].

\bibitem{Kasdagli:2023nzj}
E.~Kasdagli, M.~Ulloa and R.~P.~Woodard,
Phys. Rev. D \textbf{107}, no.10, 105023 (2023)
doi:10.1103/PhysRevD.107.105023
[arXiv:2302.04808 [gr-qc]].

\bibitem{Woodard:2023cqi}
R.~P.~Woodard and B.~Yesilyurt,
JHEP \textbf{08}, 124 (2023)
doi:10.1007/JHEP08(2023)124
[arXiv:2305.17641 [gr-qc]].

\bibitem{Miao:2024nsz}
S.~P.~Miao, N.~C.~Tsamis and R.~P.~Woodard,
Class. Quant. Grav. \textbf{41}, no.21, 215007 (2024)
doi:10.1088/1361-6382/ad7dc8
[arXiv:2405.01024 [gr-qc]].

\bibitem{Miao:2024shs}
S.~P.~Miao, N.~C.~Tsamis and R.~P.~Woodard,
[arXiv:2409.12003 [gr-qc]].

\bibitem{Arnowitt:1962hi}
R.~L.~Arnowitt, S.~Deser and C.~W.~Misner,
Gen. Rel. Grav. \textbf{40}, 1997-2027 (2008)
doi:10.1007/s10714-008-0661-1
[arXiv:gr-qc/0405109 [gr-qc]].

\bibitem{Miao:2025}
``Renormalization Group Resummation of Inflationary Graviton Loop Corrections to Massless Fermions,''
preprint in preparation.

\bibitem{Glavan:2023tet}
D.~Glavan, S.~P.~Miao, T.~Prokopec and R.~P.~Woodard,
JHEP \textbf{08}, 195 (2023)
doi:10.1007/JHEP08(2023)195
[arXiv:2307.09386 [gr-qc]].

\bibitem{Capper:1979ej}
D.~M.~Capper,
J. Phys. A \textbf{13}, 199 (1980)
doi:10.1088/0305-4470/13/1/022

\bibitem{Unruh:1998ic}
W.~Unruh,
[arXiv:astro-ph/9802323 [astro-ph]].

\bibitem{Abramo:2001dc}
L.~R.~Abramo and R.~P.~Woodard,
Phys. Rev. D \textbf{65}, 063515 (2002)
doi:10.1103/PhysRevD.65.063515
[arXiv:astro-ph/0109272 [astro-ph]].

\bibitem{Geshnizjani:2002wp}
G.~Geshnizjani and R.~Brandenberger,
Phys. Rev. D \textbf{66}, 123507 (2002)
doi:10.1103/PhysRevD.66.123507
[arXiv:gr-qc/0204074 [gr-qc]].

\bibitem{Garriga:2007zk}
J.~Garriga and T.~Tanaka,
Phys. Rev. D \textbf{77}, 024021 (2008)
doi:10.1103/PhysRevD.77.024021
[arXiv:0706.0295 [hep-th]].

\bibitem{Tsamis:2007is}
N.~C.~Tsamis and R.~P.~Woodard,
Phys. Rev. D \textbf{78}, 028501 (2008)
doi:10.1103/PhysRevD.78.028501
[arXiv:0708.2004 [hep-th]].

\bibitem{Higuchi:2011vw}
A.~Higuchi, D.~Marolf and I.~A.~Morrison,
Class. Quant. Grav. \textbf{28}, 245012 (2011)
doi:10.1088/0264-9381/28/24/245012
[arXiv:1107.2712 [hep-th]].

\bibitem{Miao:2011ng}
S.~P.~Miao, N.~C.~Tsamis and R.~P.~Woodard,
Class. Quant. Grav. \textbf{28}, 245013 (2011)
doi:10.1088/0264-9381/28/24/245013
[arXiv:1107.4733 [gr-qc]].

\bibitem{Tanaka:2012wi}
T.~Tanaka and Y.~Urakawa,
PTEP \textbf{2013}, 083E01 (2013)
doi:10.1093/ptep/ptt057
[arXiv:1209.1914 [hep-th]].

\bibitem{Tanaka:2013xe}
T.~Tanaka and Y.~Urakawa,
PTEP \textbf{2013}, no.6, 063E02 (2013)
doi:10.1093/ptep/ptt037
[arXiv:1301.3088 [hep-th]].

\bibitem{Tanaka:2013caa}
T.~Tanaka and Y.~Urakawa,
Class. Quant. Grav. \textbf{30}, 233001 (2013)
doi:10.1088/0264-9381/30/23/233001
[arXiv:1306.4461 [hep-th]].

\bibitem{Tanaka:2014ina}
T.~Tanaka and Y.~Urakawa,
PTEP \textbf{2014}, no.7, 073E01 (2014)
doi:10.1093/ptep/ptu071
[arXiv:1402.2076 [hep-th]].

\bibitem{Glavan:2016bvp}
D.~Glavan, S.~P.~Miao, T.~Prokopec and R.~P.~Woodard,
Class. Quant. Grav. \textbf{34}, no.8, 085002 (2017)
doi:10.1088/1361-6382/aa61da
[arXiv:1609.00386 [gr-qc]].

\bibitem{Miao:2017feh}
S.~P.~Miao, T.~Prokopec and R.~P.~Woodard,
Phys. Rev. D \textbf{96}, no.10, 104029 (2017)
doi:10.1103/PhysRevD.96.104029
[arXiv:1708.06239 [gr-qc]].

\bibitem{Tsamis:1989yu}
N.~C.~Tsamis and R.~P.~Woodard,
Annals Phys. \textbf{215}, 96-155 (1992)
doi:10.1016/0003-4916(92)90301-2

\bibitem{Modanese:1994wv}
G.~Modanese,
Riv. Nuovo Cim. \textbf{17N8}, 1-62 (1994)
doi:10.1007/BF02724514
[arXiv:hep-th/9410086 [hep-th]].

\bibitem{Rovelli:2001my}
C.~Rovelli,
Phys. Rev. D \textbf{65}, 044017 (2002)
doi:10.1103/PhysRevD.65.044017
[arXiv:gr-qc/0110003 [gr-qc]].

\bibitem{Giddings:2005id}
S.~B.~Giddings, D.~Marolf and J.~B.~Hartle,
Phys. Rev. D \textbf{74}, 064018 (2006)
doi:10.1103/PhysRevD.74.064018
[arXiv:hep-th/0512200 [hep-th]].

\bibitem{Giddings:2007nu}
S.~B.~Giddings and D.~Marolf,
Phys. Rev. D \textbf{76}, 064023 (2007)
doi:10.1103/PhysRevD.76.064023
[arXiv:0705.1178 [hep-th]].

\bibitem{Green:2008kj}
D.~R.~Green,
Phys. Rev. D \textbf{78}, 064066 (2008)
doi:10.1103/PhysRevD.78.064066
[arXiv:0804.4450 [hep-th]].

\bibitem{Khavkine:2011kj}
I.~Khavkine,
Phys. Rev. D \textbf{85}, 124014 (2012)
doi:10.1103/PhysRevD.85.124014
[arXiv:1111.7127 [gr-qc]].

\bibitem{Donnelly:2015hta}
W.~Donnelly and S.~B.~Giddings,
Phys. Rev. D \textbf{93}, no.2, 024030 (2016)
[erratum: Phys. Rev. D \textbf{94}, no.2, 029903 (2016)]
doi:10.1103/PhysRevD.93.024030
[arXiv:1507.07921 [hep-th]].

\bibitem{Marolf:2015jha}
D.~Marolf,
Class. Quant. Grav. \textbf{32}, no.24, 245003 (2015)
doi:10.1088/0264-9381/32/24/245003
[arXiv:1508.00939 [gr-qc]].

\bibitem{Frob:2017gyj}
M.~B.~Fr\"ob and W.~C.~C.~Lima,
Class. Quant. Grav. \textbf{35}, no.9, 095010 (2018)
doi:10.1088/1361-6382/aab427
[arXiv:1711.08470 [gr-qc]].

\bibitem{Becker:2018quq}
M.~Becker and C.~Pagani,
Phys. Rev. D \textbf{99}, no.6, 066002 (2019)
doi:10.1103/PhysRevD.99.066002
[arXiv:1810.11816 [gr-qc]].

\bibitem{Becker:2019tlf}
M.~Becker and C.~Pagani,
Universe \textbf{5}, no.3, 75 (2019)
doi:10.3390/universe5030075

\bibitem{Donoghue:1996mt}
J.~F.~Donoghue and T.~Torma,
Phys. Rev. D \textbf{54}, 4963-4972 (1996)
doi:10.1103/PhysRevD.54.4963
[arXiv:hep-th/9602121 [hep-th]].

\bibitem{Glavan:2024elz}
D.~Glavan, S.~P.~Miao, T.~Prokopec and R.~P.~Woodard,
JHEP \textbf{03}, 129 (2024)
doi:10.1007/JHEP03(2024)129
[arXiv:2402.05452 [hep-th]].

\bibitem{Katuwal:2021thy}
S.~Katuwal and R.~P.~Woodard,
JHEP \textbf{10}, 029 (2021)
doi:10.1007/JHEP10(2021)029
[arXiv:2107.13341 [gr-qc]].

\bibitem{Bjerrum-Bohr:2002aqa}
N.~E.~J.~Bjerrum-Bohr,
Phys. Rev. D \textbf{66}, 084023 (2002)
doi:10.1103/PhysRevD.66.084023
[arXiv:hep-th/0206236 [hep-th]].

\bibitem{Glavan:2019msf}
D.~Glavan, S.~P.~Miao, T.~Prokopec and R.~P.~Woodard,
JHEP \textbf{10}, 096 (2019)
doi:10.1007/JHEP10(2019)096
[arXiv:1908.06064 [gr-qc]].

\bibitem{Miao:2018bol}
S.~P.~Miao, T.~Prokopec and R.~P.~Woodard,
Phys. Rev. D \textbf{98}, no.2, 025022 (2018)
doi:10.1103/PhysRevD.98.025022
[arXiv:1806.00742 [gr-qc]].

\bibitem{Miao:2008sp}
S.~P.~Miao and R.~P.~Woodard,
Class. Quant. Grav. \textbf{25}, 145009 (2008)
doi:10.1088/0264-9381/25/14/145009
[arXiv:0803.2377 [gr-qc]].

\bibitem{Woodard:2008yt}
R.~P.~Woodard,
Phys. Rev. Lett. \textbf{101}, 081301 (2008)
doi:10.1103/PhysRevLett.101.081301
[arXiv:0805.3089 [gr-qc]].

\end{thebibliography}
\end{document}